\documentclass[aps,pra,reprint,superscriptaddress,floatfix,nofootinbib]{revtex4-2}

\usepackage[english]{babel}
\usepackage[utf8]{inputenc}
\usepackage[T1]{fontenc}

\usepackage{amsfonts,amsmath,amssymb,accents,braket,mathtools,derivative}
\usepackage{dsfont,amsfonts,bm}
\usepackage{enumitem}

\usepackage{placeins}
\usepackage{graphicx}
\graphicspath{ {figures/},{figures/NB_5_NI_1/},{figures/NB_2_NI_2/},{figures/NB_5_NI_2/},{figures/NB_10_NI_1/}}
\usepackage{epstopdf}

\usepackage[colorlinks=true,citecolor=blue,linkcolor=blue,urlcolor=blue]{hyperref}
\usepackage[capitalise]{cleveref}
\usepackage{tikz}
\usetikzlibrary{plotmarks}

\newcommand{\maj}{A}
\newcommand{\imp}{B}
\DeclarePairedDelimiter{\norm}{\lVert}{\rVert} 

\begin{document}

\title{Breathing dynamics of the Bose polaron in a species-selective harmonic trap}

\author{Maxim Pyzh}
\email{mpyzh@physnet.uni-hamburg.de}
\affiliation{Zentrum f\"ur Optische Quantentechnologien, Universit\"at
	Hamburg, Luruper Chaussee 149, 22761 Hamburg, Germany}
\author{Peter Schmelcher}
\email{pschmelc@physnet.uni-hamburg.de}
\affiliation{Zentrum f\"ur Optische Quantentechnologien, Universit\"at 
	Hamburg, Luruper Chaussee 149, 22761 Hamburg, Germany}
\affiliation{The Hamburg Centre for Ultrafast Imaging, Universit\"at 
	Hamburg, Luruper Chaussee 149, 22761 Hamburg, Germany}

\begin{abstract}
	We perform an extensive numerical study on the breathing dynamics
	of a few-body Bose polaron setup in a one-dimensional 
	species-selective harmonic trap.
	The dynamics is triggered by a quench of the impurity trap.
	The excitation of the background majority atoms is mediated via
	the majority-impurity interaction.
	The breathing spectrum is obtained for different numbers of majority particles, 
	several values of the majority-component interaction strengths and trap ratios.
	It is further compared to the breathing spectrum of a particle-balanced 
	few-body Bose-Bose mixture.
	In particular, for equal post-quench traps 
	the employed protocol allows to couple
	states of different center-of-mass parity
	in contrast to species-symmetric trap quenches.
	Among the participating eigenstates we identify one
	having odd center-of-mass parity and
	even global parity.
	The breathing frequency induced by this state
	is a monotonically decreasing function
	of the coupling parameter.
	Importantly, in order to be observable
	it requires the entanglement between the species to be taken into account.
	We demonstrate this by comparing the numerically exact results obtained by means of the Multi-Layer Multi-Configuration Time-Dependent Hartree Method for Mixtures
	to the ones of a species mean-field ansatz.
	The entanglement-sensitive breathing frequency persists also for unequal post-quench
	traps where the center-of-mass cannot be decoupled.
	Finally, we analyze the impact of parity symmetry on the breathing dynamics by
	initializing a state of odd global parity.
	We evidence a striking resemblance to the ground state breathing dynamics.
\end{abstract}

\maketitle

\section{Introduction}
\label{sec:intro}

The polaron concept was introduced quite some time ago by Landau
and Pekar \cite{landau1933electron,pekar1946local} 
to describe the motion of an
electron in a crystalline material. 
The notion of an emerging quasi-particle
dressed by low-energy excitations of the underlying medium
has vastly expanded since its foundation finding broad applications 
in different areas of physics such as
organic semiconductors, polymers, nanowires, 
quantum dots and high temperature superconductors
\cite{devreese2009frohlich,alexandrov2010advances,mahan2013many}.
Since the advent of ultracold gases 
\cite{anderson1995observation,davis1995bose}
a promising experimental platform has emerged allowing 
to investigate fundamental many-body quantum processes \cite{bloch2008many} 
with an exquisite tunability
of the underlying interactions and trapping geometries.
In particular, the ability to combine different species
\cite{myatt1997production}
and the precise control over the number of particles
\cite{blume2012few} made it possible
to experimentally prepare an impurity 
in a many-body environment of bosons
\cite{palzer2009quantum,catani2012quantum,fukuhara2013quantum,
	spethmann2012dynamics,jorgensen2016observation,meinert2017bloch}
or fermions 
\cite{nascimbene2009collective,schirotzek2009observation,
	koschorreck2012attractive,
kohstall2012metastability,cetina2015decoherence} 
leading to what is nowadays termed Bose \cite{grusdt2015new} 
and Fermi polaron \cite{chevy2010ultra,massignan2014polarons}, 
respectively.
A mapping to the Fröhlich Hamiltonian \cite{frohlich1954electrons} 
for the polaron problem can be established
while all the Hamiltonian parameters can be 
addressed individually.
The tunability of interactions via Feshbach resonances 
\cite{chin2010feshbach,kohler2006production}
provides access to highly correlated and entangled regimes 
challenging the theorists 
to go beyond the weak coupling Fröhlich paradigm
\cite{volosniev2017analytical,
	grusdt2017bose,grusdt2018strong,kain2018analytical,
	drescher2019real,jager2020strong,drescher2020theory}.

The correlations are in particular enhanced in quasi-one-dimensional (1D) 
systems \cite{giamarchi2003quantum}.
A comparatively tight transverse confinement 
freezes the perpendicular motion of particles and additionally
affects the effective 1D interactions 
known as confinement-induced resonances
\cite{olshanii1998atomic,bergeman2003atom,haller2010confinement}.
A prominent example of a strongly correlated 1D system 
is the Tonks-Girardeau gas
\cite{girardeau1960relationship,
	kinoshita2004observation,paredes2004tonks}.
In contrast to higher dimensions, 
where a lower particle density implies
weaker correlations, in 1D lower densities lead to stronger interactions.
It makes the study 
of low-density few-body systems of particular interest 
triggering significant research efforts 
\cite{sowinski2019one,mistakidis2019quench,mistakidis2019dissipative,
	mistakidis2019effective,mistakidis2020many}. 
At the same time this represents a great challenge
requiring sophisticated numerical techniques able to account for 
all the relevant correlations when characterizing
the static properties or the many-body dynamics, such as the 
Density Matrix Renormalization Group (DMRG) \cite{schollwock2005density} or the 
Multi-Layer Multi-Configuration Time-Dependent Hartree Method
for Bosons and Fermions
(ML-MCTDHX) \cite{cao2017unified}.
In species-selective trapping geometries \cite{catani2012quantum,bentine2017species,barker2020realising} 
the inhomogeneity of the medium and the localization length of the impurity 
impact significantly the degree of correlations \cite{keiler2019interaction,pyzh2020phase,pyzh2021entangling} opening
interesting perspectives but requiring also new approaches
since the translation symmetry is broken 
making the well established technique, 
the Lee-Low-Pines transformation \cite{lee1953motion}, inapplicable.

In this work we investigate
the low-energy breathing dynamics
of a 1D few-body Bose polaron in a species-selective parabolic confinement.
Elementary collective excitations
\cite{catani2012quantum,johnson2012breathing,moritz2003exciting,
	stoferle2004transition,nascimbene2009collective,haller2009realization,
	fang2014quench} 
are of fundamental importance
to understand the dynamical response of a physical system
subject to a weak perturbation
in terms of excited eigenstates and the respective eigenenergies.
Here, we focus on the so-called breathing modes \cite{abraham2014quantum}.
On account of their strong sensitivity to the system’s parameters such as 
interactions, trap geometry and spin statistics,
they have been established as a reliable diagnostic tool
to access the ground state properties of a system \cite{mcdonald2013theory},
for precision measurements of the scattering lengths \cite{moritz2003exciting}
and even as a sensitive test of the equation of state at unitarity \cite{astrakharchik2005equation,altmeyer2007precision}.
We make use of the species-selective trapping potentials
to compress only the trap of the impurity and monitor the response
in the majority component owing to the inter-component coupling.
For different system parameters 
we classify the breathing modes according to their relative amplitude
in the Fourier power spectrum obtained by applying
a Compressed Sensing (CS) \cite{andrade2012application} algorithm to breathing observables 
and study the role of entanglement and parity symmetry on breathing response.
We identify a unique mode
whose frequency is a monotonically 
decreasing function of the majority-impurity interaction
while its presence relies on the entanglement between
the impurity and the majority component.
 
This work is structured as follows. 
In \cref{sec:general_setup} we introduce our setup and Hamiltonian.
The numerical approach is discussed \cref{subsec:mlx}. We use ML-MCTDHX
for state initialization, subsequent dynamics
and evaluation of breathing observables.
The oscillation frequencies are extracted 
by means of a CS algorithm outlined in \cref{subsec:compressed_sensing}.
The results presented in \cref{sec:breathing} are 
categorized in four subtopics: 
an overview of breathing modes in a particle-balanced 
few-body Bose-Bose mixture \cite{pyzh2018spectral} (\cref{subsec:bose:mixture}) 
for later reference,
the breathing spectra in the current Bose polaron setup
for different majority-component interactions and particle number ratio 
(\cref{subsec:polaron_breathing}),
the impact of impurity localization length (\cref{subsec:inhomogeneity})
and, finally, the role of global parity symmetry for the breathing response
(\cref{subsec:1excbreathing}).
In \cref{sec:conclusions} we summarize the most important insights.


\section{Setup and Hamiltonian}
\label{sec:general_setup}

We consider a few-body mixture of two bosonic components.
A component $\sigma \in \{\maj, \imp\}$
contains $N_{\sigma}$ particles of mass $m_{\sigma}$,
which experience a quasi one-dimensional parabolic confinement 
with trap frequency $\omega_{\sigma}$
and interact internally
via contact pseudo interaction of strength $g_{\sigma}$.
The components are coupled via an inter-species contact interaction 
of strength $g_{\maj \imp}$.
We assume equal masses and introduce harmonic units 
of component $\maj$ as our natural units, i.e., 
$l_{\maj} = \sqrt{\hbar/ m \omega_{\maj}}$ for length, 
$\hbar \omega_{\maj}$ for energy
and $1/\omega_{\maj}$ for time.
The corresponding Hamiltonian reads as follows:
\begin{align}
	H & = H_{\maj} + H_{\imp} + H_{\maj \imp} =
	\label{eq:hamiltonian}\\
	& = \sum_{i=1}^{N_{\maj}} \left( -\frac{1}{2} 
	\frac{\partial^2}{\partial x_i^2} 
	+ \frac{1}{2} x_i^2 \right) +
	g_{\maj} \sum_{i<j}^{N_{\maj}} \delta(x_i-x_j) 
	+ \nonumber\\
	& + \sum_{i=1}^{N_{\imp}} \left( -\frac{1}{2} 
	\frac{\partial^2}{\partial y_i^2} + 
	\frac{1}{2} \eta^2 y_i^2 \right) +
	g_{\imp} \sum_{i<j}^{N_{\imp}} \delta(y_i-y_j)  
	+ \nonumber \\
	& + g_{\maj \imp} \sum_{i=1}^{N_{\maj}} 
	\sum_{j=1}^{N_{\imp}} \delta(x_i-y_j), 
	\nonumber
\end{align}
where $x_i$ ($y_i$) denotes the position 
of the $i$th particle in the $A$ ($B$) component.

The $\maj$ component is referred to as the majority species.
It has $N_{\maj} \in \{5,10\}$ particles.
The $\imp$ component consists of a single particle, $N_{\imp}=1$,
and we call it the impurity.
The majority-component is either non-interacting ($g_{\maj}=0$)
or features a weakly attractive/repulsive interaction ($g_{\maj}=\pm 0.5$).
The majority-impurity coupling covers values from weakly attractive
to intermediate repulsive $g_{\maj \imp} \in [-0.5,2.0]$.
The trap ratio covers cases of equal traps ($\eta=1$),
a 'broad' impurity ($\eta=0.51$) and a 'narrow' impurity ($\eta=4$).

\begin{figure}[h]
	\centering
	\includegraphics[width=1\columnwidth]{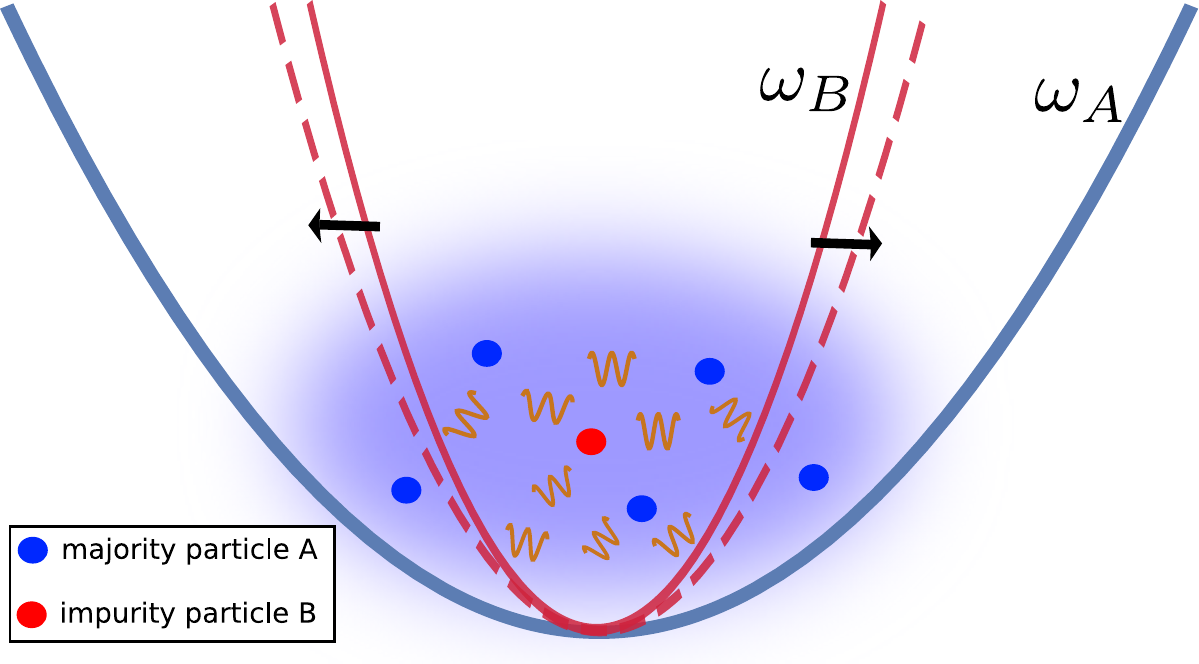}
	\caption{The impurity (species B indicated by a red circle) in a harmonic trap 
		(red solid line) is immersed in a cloud of majority atoms 
		(species A indicated by a blue ellipse) subject to a different parabolic confinement 
		(blue solid line).
		The breathing dynamics is initiated 
		by quenching the trap of the impurity
		(red dashed line) inducing thereby excitations (orange waves) 
		in the composite system
		via the majority-impurity interaction.}
	\label{fig:setup}
\end{figure}

The Hamiltonian \cref{eq:hamiltonian} 
possesses a global reflection symmetry, corresponding to a map
$x_i \mapsto -x_i$ and $y_j \mapsto -y_j$ for all $i,j$.
The eigenstates can be thus partitioned into
two decoupled subspaces of even and odd global parity.
As our initial state we take the lowest energy
eigenstate from either of the two subspaces.
To initiate the breathing dynamics we slightly relax the
trap of the impurity from $1.05\eta$ to $\eta$.
The majority species is set in breathing motion indirectly 
via the coupling to the impurity.

The breathing motion can be monitored 
in the one-body densities $\rho_1^{\sigma}(z,t)$
as their widths expand and contract periodically in time.
Alternatively, one can analyze the time-evolution of 
the corresponding breathing observables $\braket{\hat{O}}_t$
with $\hat{O} = \sum_i \hat{x}_i^2$ for the majority species 
and $\hat{O}=\hat{y}^2$ for the impurity.
Such an oscillatory motion is usually
composed of multiple contributions
with distinct amplitudes $d_n$ and frequencies $\Omega_n$, i.e.,
$\braket{\hat{O}}_t=\sum_n d_n e^{-i \Omega_n t}$.
Each of the oscillatory contributions in $\braket{\hat{O}}_t$ will be
referred to as a {\it breathing mode} 
characterized by a distinct frequency.
The origin of the breathing modes can be expressed in terms
of system's eigenstates and related eigenenergies as follows.
Upon a quench, several eigenstates $\ket{\psi_n}$ 
of the post-quench Hamiltonian $H$ with energy $E_n$
become populated depending on the overlap $c_n$
with the initial state $\ket{\Psi(t_0)}$, 
i.e., $c_n = \braket{\psi_n | \Psi(t_0)}$.
The time signal $\braket{\hat{O}}_t=
\sum_{n,m} c^*_m c_n \bra{\psi_m}\hat{O}\ket{\psi_n} 
e^{-i \Omega_{n,m} t}$ oscillates with 
frequencies $\Omega_{n,m}=E_n-E_m$
for populated eigenstates $c_n c_m \neq 0$
as long as the transition matrix elements 
$\braket{\psi_n|\hat{O}|\psi_m} = O_{nm}$ are non-zero.
Given a weak perturbation,
we expect the lowest energy eigenstate
of even (odd) parity $\ket{\psi_{\rm{ref}}}$ to have the largest overlap
with the even (odd) parity initial state $\ket{\Psi(t_0)}$.
For that reason, the major part of frequencies
contained in the breathing observable $\braket{\hat{O}}_t$
are energy gaps between any of the populated eigenstates and the reference eigenstate,
i.e., $\Omega_{n}=E_n-E_{{\rm ref}}$.
Oscillations among any different combination of eigenstates are of minor amplitude,
so we don't focus on them by setting a suitable amplitude threshold.
For that reason, when we refer to an eigenstate of a breathing mode
we mean the eigenstate responsible for this mode
while the reference state $\ket{E_{\rm ref}}$
is usually clear within the context, unless we state otherwise. 

For $\eta=1$ the center-of-mass coordinate can be decoupled
by employing a transformation 
to the center-of-mass and internal coordinates such as Jacobi coordinates.
This provides an exact quantum number 
of the 'free' center-of-mass harmonic oscillator.
At first glance, it is appealing to operate in the center-of-mass frame.
However, the two-body interactions are mapped to higher order interaction terms
while the exchange symmetry for indistinguishable majority coordinates 
in the laboratory frame becomes a set of complex rules in the center-of-mass frame.
Furthermore, for $\eta \neq 1$ the center-of-mass cannot be separated.
Hence, we employ the laboratory frame.

\section{Computational Approach and Analysis}
\label{sec:methodology}

In this work we use the Multi-Layer Multi-Configuration 
Time-Dependent Hartree Method for Mixtures
to initialize a system in its ground state by means of relaxation,
i.e., propagation in imaginary time, for the state evolution
following a trap quench of the impurity and to evaluate the expectation values
of breathing observables for each species as a function of time.
We outline the major idea of the method in \cref{subsec:mlx} along with
the wave function ansatz for the system at hand.
We then apply a Compressed Sensing 
algorithm to retain the frequencies from those observables.
In view of the fact that CS relies on the sparsity condition
in the Fourier space, it is not used as a standard tool
for frequency extraction from a time signal.
In \cref{subsec:compressed_sensing} we discuss advantages 
of this method in the current application as compared to a straightforward Fourier
transformation.
In order to be self-contained we also provide the implementation details.

\subsection{ML-MCTDHX}
\label{subsec:mlx}

To prepare the initial state $\ket{\Psi(t_0)}$ 
and to perform the subsequent time evolution 
$\ket{\Psi(t)}=e^{-iHt}\ket{\Psi(t_0)}$ 
with the time-independent Hamiltonian $H$ from \cref{eq:hamiltonian}
we employ the 
{\it Multi-Layer Multi-Configuration 
Time-Dependent Hartree Method 
for Mixtures} of indistinguishable particles, 
for short ML-X \cite{cao2013multi,kronke2013non,cao2017unified}.
The core idea behind ML-X is 
to expand the many-body wave-function 
in a properly symmetrized product state basis,
the so-called Fock states, such that
the underlying single-particle functions (SPFs) are
{\it time-dependent}.
These are variationally optimized
during the time evolution 
to provide a more 'compact' description
compared to Fock states composed of time-independent SPFs.
Compact means that in general much less SPFs are required
reducing thus the Fock space dimension
while retaining a similar degree of accuracy.

As the system evolves, 
the many-body state travels through different 
subspaces of the complete Hilbert space.
If the Fock basis is fixed, in general 
a large set of SPFs is required 
to cover all the relevant subspaces.
Many configurations become 'actively' populated 
during the time-evolution,
though not necessarily all of them 
at the same time with a fraction staying or becoming inactive.
Even when all of Fock states are populated,
we may rotate the basis by choosing a different set of SPFs
until, eventually, we end up with a more compact representation.
Given a truncated Fock space, ML-X rotates the basis vectors
such as to find the best possible representation of
the exact many-body state at each instant of time.
In other words, it looks for the current 'active' subspace.
Once the truncated
variationally evolving Fock-space
becomes large enough to contain the (major part of) 
'active' subspace,
the representation of $\ket{\Psi(t)}$ given by ML-X 
is considered optimal.

The underlying wave function ansatz for the Bose polaron problem 
belonging to \cref{eq:hamiltonian}
is expanded in two layers (Multi-Layer):

\begin{align}
	\ket{\Psi(t)} 
	&=\sum_{i=1}^{S}\sqrt{\lambda_i(t)} 
	\ket{\Psi_i^{\maj}(t)} \otimes \ket{\Psi_i^{\imp}(t)},
	\label{eq:wfn_ansatz_species_layer} \\
	\ket{\Psi_i^{\sigma}(t)} 
	&=\sum_{\vec{n}^{\sigma}|N_\sigma}
	C_{i, \vec{n}^{\sigma}}(t)
	\ket{\vec{n}^{\sigma}(t)}.
	\label{eq:wfn_ansatz_particle_layer}
\end{align}

In the first step, see \cref{eq:wfn_ansatz_species_layer},
the majority and impurity degrees of freedom, $x_i$ and $y$ respectively, 
are separated
and assigned to $S \in \mathbb{N}$ time-dependent 
species wave-functions $\ket{\Psi_i^{\sigma}(t)}$.
The sum of product form is very convenient as it makes evident
the entanglement between the two components.
The time-dependent weights $\lambda_i(t)$ are normalized 
$\sum_i^S \lambda_i(t)=1$
and sorted in descending order.
A composite system with $\lambda_1(t) \approx 1$ is considered disentangled.
Assuming $S=1$ in the expansion
is called a species mean-field approximation.
In the second step, see \cref{eq:wfn_ansatz_particle_layer},
the species wave-functions belonging to the same component 
are expanded in
the same Fock-state basis $\ket{\vec{n}^{\sigma}(t)}$ 
with time-dependent coefficients $C_{i, \vec{n}^{\sigma}}(t)$.
The time-dependence of number states is meant implicitly 
through the time-dependence
of $s_{\sigma}\in \mathbb{N}$ underlying SPFs $\varphi_j^{\sigma}(t)$
which are represented using a harmonic discrete variable representation (DVR)
\cite{light1985generalized}.
The notation $\vec{n}^{\sigma}|N_{\sigma}$ denotes particle number conservation,
i.e., $\sum_i^{s_{\sigma}} n_i^{\sigma} = N_{\sigma}$.
Finally, by applying the Dirac-Frenkel variational principle \cite{raab2000dirac}
the equations of motion for $\lambda_i$, $C_{i, \vec{n}^{\sigma}}$
and $\varphi_j^{\sigma}$ are obtained.
The convergence of ML-X is controlled via $S$, $s_{\sigma}$ and the number
of DVR grid points. We use $S=s_{\sigma}=8$ for $N_A=5$ and
$S=s_{\sigma}=6$ for $N_A=10$. The DVR grid spans an interval $[-6,6]$
and we choose $151$ DVR grid points.

\subsection{Compressed Sensing Analysis}
\label{subsec:compressed_sensing}

In this work, we aim to extract frequencies 
$\Omega_{n,m}=E_n-E_m$
of system's excitations where $E_n$
denotes the eigenenergy of the
$n$-th eigenvector $\ket{\psi_n}$ of $H$.
Any physical observable $\hat{O}$ carries information
about excited eigenstates
$\braket{\hat{O}}_t=
\sum_{n,m} c^*_m c_n \bra{\psi_m}\hat{O}\ket{\psi_n} 
e^{-i \Omega_{n,m} t}$,
as long as the transition matrix elements 
$\bra{\psi_m}\hat{O}\ket{\psi_n}$
are non-zero and the corresponding eigenstates 
are initially populated $c_n c_m \neq 0$ 
where $c_n=\braket{\psi_n|\Psi(t_0)}$.
We perform a sampling of breathing observables 
with a uniform rate $\Delta t$
over an interval $[0,T]$ containing $T/\Delta t+1=N_t$ points.
It gives us a finite time signal ${\bf f} \in \mathbb{R}^{N_t}$ 
with components $f_j$ 
of discrete variable $t_j \in \mathbb{R}$, i.e.,
$f_j = f(t_j) = f(\Delta t \cdotp j)$ 
with integer index $j \in [0,N_t-1] \subset \mathbb{N}_0$.

A straightforward way to retain the frequencies contained
in ${\bf f}$ is to perform a
discrete Fourier transformation (DFT),
expressed as a linear map ${\bf A}{\bf f}={\bf g}$ 
with a square matrix ${\bf A} \in \mathbb{C}^{N_t \times N_t}$ 
and signal's representation in the frequency domain
${\bf g} \in \mathbb{C}^{N_t}$.
The latter is characterized by frequency spacing 
$\Delta \omega$ and cut-off frequency $\omega_{cut}$,
i.e., it has components 
$g_j= g(\omega_j) = g(\Delta \omega \cdotp j)$
of discrete variable $\omega_j \in \mathbb{R}$
and for odd (even) $N_{\omega}=N_t$ number of points
spans an open (closed) interval 
with endpoints $-\omega_{cut}/2$ and $\omega_{cut}/2$.

The sampling parameters of time and frequency domain
are interrelated. Thus, the sampling time $T$ determines
the frequency spacing $\Delta \omega = 2\pi/T$, 
while the sampling rate $\Delta t$ determines the 
Nyquist frequency $\omega_{cut}=\pi/\Delta t$.
In principle, frequencies can be retained with arbitrary resolution,
if sampled long enough, while highly oscillatory components require
a finer sampling rate.
In practice, there are technical limitations 
such as generation, storage and processing of data.
Given a complex system such as ours, 
data generation becomes a time-consuming factor
making a good resolution in frequency domain out of reach.

In order to overcome this obstacle, 
some prior information about signal's properties might become useful.
Since the system is perturbed weakly,
we expect only the low-energy excitations 
to be of relevance for the underlying dynamics.
In particular, we expect ${\bf g}$
to be sparse with major components located 
in the low-frequency region.

\begin{figure*}[t]
	\centering
	\includegraphics[width=1.0\textwidth]{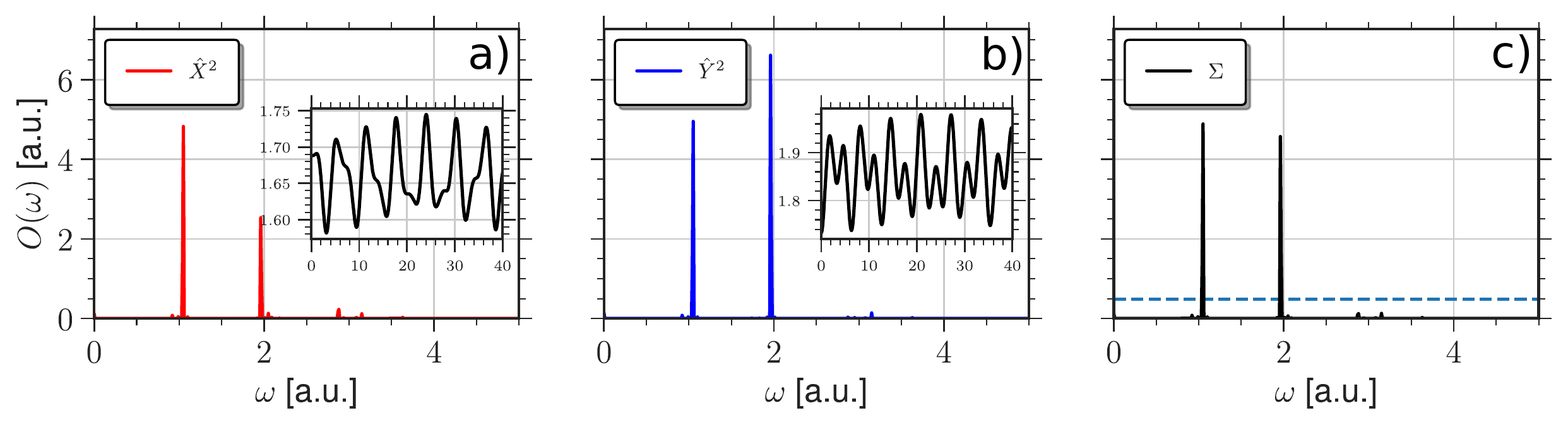}
	\caption
	{Fourier power spectrum $X^2(\omega)$ (a) and $Y^2(\omega)$ (b)
		obtained
		by applying a Compressed Sensing algorithm to the expectation values of
		the breathing observables
		$\braket{\sum_i \hat{x}_i^2}_t$ and
		$\braket{\hat{y}^2}_t$ (insets) evaluated w.r.t. a dynamical state
		$\ket{\Psi(t)}$ obtained by the Multi-Layer Multi-Configuration 
		Time-Dependent Hartree Method for Mixtures.
		In the averaged power spectrum $\Sigma(\omega)$ (c) the blue dashed line
		indicates a threshold magnitude and only frequencies above it are accounted for
		in \cref{sec:breathing} (see text).
		The physical parameters are $\eta=1$, $N_{\maj}=5$, $N_{\imp}=1$, 
		$g_{\maj \imp}=2.0$, $g_{\maj}=0$ (see \cref{sec:general_setup})
		and the Compressing Sensing parameters are $T=40$, $\Delta t=0.1$, 
		$\omega_{cut}=20$ and $\Delta\omega=0.01$ (see \cref{subsec:compressed_sensing}).}
	\label{fig:power_spectrum}
\end{figure*}

With this prior knowledge {\it compressed sensing} (CS)
allows to retain the frequencies with a very high resolution
while keeping the simulation time with ML-X reasonably small.
To this end, we formulate our problem
as finding the vector ${\bf g}$ satisfying the inverse DFT condition,
i.e., ${\bf f = A^\dagger g}$.
However, ${\bf A^\dagger} \in \mathbb{C}^{N_t \times N_w}$ 
is now a rectangular matrix with $N_w \gg N_t$ and 
${\bf g} \in \mathbb{C}^{N_{\omega}}$ 
resulting in an underdetermined system of equations.
Here, ${\bf A^\dagger}$ is a submatrix
of the inverse square DFT matrix 
${\bf A^\dagger}\in \mathbb{C}^{N_{\omega} \times N_{\omega}}$,
with the last $N_{\omega}-N_t$ rows being removed.
Importantly, the number of columns $N_{\omega}$ 
and thus the frequency spacing
$\Delta \omega$ can be chosen independently of the simulation time $T$.
Intuitively, this implies that ${\bf g}$ has been generated by 
a signal extended over a larger region $T'>T$ 
than the current one ${\bf f}$,
though the information contained beyond $T$ is considered redundant
given the priors underlying the evolution.

In order to find a sparse solution 
to a linear ill-posed inverse problem
we formulate it as $\ell1$-norm penalized 
least-squares minimization task
known as {\it basis pursuit denoising} (BPDN) \cite{van2009probing}:
\begin{equation}
	\min_{\bf g} \frac{1}{2} \norm{ {\bf f}-{\bf A^{\dagger} g} }_2^2 
	+ \lambda \norm{{\bf g}}_1,
	\label{eq:l1_optimization}	
\end{equation}
where $\norm{ {\bf x} }_p=\left(\sum_i |x_i|^p\right)^{1/p}$ 
while the penalty term $\lambda \geq 0$ controls the trade-off 
between the sparsity of the solution and the constraint violation
in the presence of noisy signal ${\bf f}$.
We use the Least Angle Regression (LARS) \cite{loris2008l1packv2} minimization algorithm 
to solve \cref{eq:l1_optimization} and perform a mean-normalization 
of the signal ${\bf f} \rightarrow 
{\bf \tilde{f}} = ({\bf f} - {\bf \overline{f}}) / \norm{ {\bf f} }_1$
beforehand.
The employed implementation 
requires real inputs in \cref{eq:l1_optimization}.
Real ${\bf f}$ implies hermitian ${\bf g}$ and we use this symmetry
to reformulate ${\bf g}$ 
as a real vector: 
${\bf g} \rightarrow {\bf \tilde{g}} \in \mathbb{R}^{2 N_{\omega}} 
= (\operatorname{Re}({\bf g}),\operatorname{Im}({\bf g}))$.
Correspondingly, the inverse Fourier matrix 
${\bf A}^{\dagger} \rightarrow
{\bf M} \in \mathbb{R}^{N_t \times 2 N_{\omega}} = ({\bf C}, {\bf D})$ 
is now composed of two real sub-matrices
${\bf C}\in \mathbb{R}^{N_t \times N_{\omega}}$ with components
$c_{i,j}=\cos(\epsilon i j)$ and
${\bf D}\in \mathbb{R}^{N_t \times N_{\omega}}$ with components
$d_{i,j}=\sin(\epsilon i j)$ where $\epsilon=\Delta t \Delta \omega$.
The ML-X time-domain parameters are chosen as 
$T=40$ and $\Delta t=0.05$, 
whereas the CS frequency-domain parameters are
$\omega_{cut}=20$ and 
$\Delta \omega=0.01$.\footnote{We remark that similar resolution 
with DFT can be obtained with $T \approx 600$.}

As input time signals ${\bf f}$ we use the expectation values of breathing observables 
$\sum_i \hat{x}_i^2$ for the majority species
and $\hat{y}^2$ for the impurity evaluated 
w.r.t.\ the dynamical state $\ket{\Psi(t)}$
obtained by ML-X.
We apply the CS algorithm to obtain the corresponding 
vector ${\bf \tilde{g}}$.
Then, we map ${\bf \tilde{g}}$ back to ${\bf g}$ and 
convert complex values into amplitudes, i.e., $g_j \rightarrow |Re(g_j)+Im(g_j)|$.
The final vector we call a Fourier power spectrum and
label it as $X^2(\omega)$ for the majority component and $Y^2(\omega)$ for the impurity.
Finally, for a fixed set of physical parameter values 
we construct an averaged power spectrum $\Sigma=(X^2+Y^2)/2$.
An example is shown in \cref{fig:power_spectrum}.
Each frequency is classified as being of
a majority type (red), an impurity type (blue) or of a mixed type
depending on the relative weights of $X^2$ and $Y^2$ in $\Sigma$.
These are encoded in subsequent figures as a pie chart of two colors 
for each breathing frequency.
Additionally, we use the transparency 
to indicate the magnitude of participating breathing modes
relative to the most relevant mode of amplitude $\max(\Sigma)=\Sigma_{\rm max}$
such that faded colors imply less relevant modes.
Frequencies with a contribution below $\Sigma_{\rm max}/10$ 
are discarded. This amounts to neglecting
i) low-amplitude oscillations among any two eigenstates not involving 
the reference eigenstate with the largest population,
and ii) numerically introduced 'phantom' peaks 
which are also of minor amplitude.

\section{Results}
\label{sec:breathing}

First, in \cref{subsec:bose:mixture} 
we summarize results concerning the breathing dynamics
of a single particle, a single-component condensate, 
two distinguishable particles and a particle-balanced Bose-Bose mixture.
This will provide us with useful insights for the interpretation of breathing modes
unraveled in the Bose polaron setup being the subject of \cref{subsec:polaron_breathing}.
In \cref{subsec:inhomogeneity} we investigate the impact of the 
trap ratio $\eta$ on the breathing spectrum accounting for two cases:
a 'broad' ($\eta<1$) and a 'narrow' ($\eta>1$) impurity.
Finally, in \cref{subsec:1excbreathing} we study the breathing
response of the first excited state having odd parity
and contrast it to the response of the ground state 
which is of even parity.

In the following, when all interactions are zero  
we employ a notation
$\ket{\vec{n}}=\ket{n_1, n_2, \ldots}$ 
to denote $n_i$ particles occupying the $i$-th orbital
of a single-particle quantum harmonic oscillator.
It is not to be confused with permanents introduced in \cref{subsec:mlx}
where the orbitals are variationally optimal at each time instant.
We also drop the redundant zeroes in the vector tail 
once all the particles have been accounted for,
i.e., $\sum_i n_i = N$.
The notation $\ket{\vec{n}^{A}} \otimes \ket{\vec{n}^{B}}$ denotes a product state
for the two components. 
Note that for unequal traps ($\eta \neq 1$)
the orbitals for each species are different.

\subsection{Bose-Bose mixture}
\label{subsec:bose:mixture}

The breathing mode frequency of a single particle confined 
in a parabolic trap of frequency $\omega$
is known to be $\Omega = 2 \omega$, 
corresponding to an excitation 
by two energy quanta $\ket{0,0,1}$
w.r.t.\ the harmonic oscillator basis.
An ensemble of $N$ non-interacting ($g=0$) bosonic particles 
introduces an additional eigenstate 
of the same excitation energy $2 \omega$,
namely a two-particle excitation $\ket{N-2,2}$ 
being degenerate with $\ket{N-1,0,1}$.
For interacting particles the degeneracy is lifted.
The frequency of one mode remains constant for any $g$ 
and relates to the center-of-mass (CM) breathing motion. 
The frequency of the other mode 
is highly sensitive to a variation of $g$
and characterizes the relative motion of particles \cite{schmitz2013quantum}.
A mean-field ansatz for the breathing dynamics, 
being a single-particle picture,
is able to recover only the interaction-sensitive breathing frequency,
though with quantitative deviations 
as compared to an exact solution, 
especially at sizable interactions.

Two distinguishable non-interacting particles $A$ and $B$
confined in the same harmonic trap ($\eta=1$)
have three eigenstates with excitation energy $2 \omega$:
a single particle excitation by two energy quanta 
in either of the components, i.e.,
$\ket{0,0,1} \otimes \ket{1}$ and $\ket{1} \otimes \ket{0,0,1}$, 
as well as a two-particle-excitation $\ket{0,1} \otimes \ket{0,1}$.
The latter can be also interpreted as a simultaneous excitation 
of a sloshing mode in each species.
When the coupling between the particles becomes non-zero,
the degenerate manifold splits giving rise to
one constant frequency mode ($\Omega=2\omega$) and two modes
with an interaction sensitive oscillation period (not shown).
One of the latter two modes features a frequency
which is a convex function of $g_{AB}$.
The other mode is remarkable
in the sense that the energy
of the involved excited eigenstate is independent of the interaction
while the ground state energy (reference eigenstate) 
grows with increasing $g_{AB}$.
As a result, the corresponding breathing mode frequency
is a monotonically decreasing function
of the coupling $g_{AB}$.
It saturates to a sloshing mode frequency $\Omega=\omega$
at very large positive couplings.
Thus, distinguishability allows for
an additional breathing mode.
A species mean-field ansatz 
can identify
two coupling sensitive breathing frequencies
though neither of the exact modes 
will be matched quantitatively
for all interactions.
In particular, the monotonically decreasing frequency
is not accounted for
implying the relevance of entanglement 
in multi-component mixtures.

A non-interacting two-component mixture 
features in total five eigenstates 
which are two energy quanta above the ground state:
two single-particle excitation states 
$\ket{N_A-1,0,1} \otimes\ket{N_B}$ and 
$\ket{N_A} \otimes \ket{N_B-1,0,1}$,
two states having two indistinguishable particles excited
$\ket{N_A-2,2} \otimes\ket{N_B}$ and $\ket{N_A} \otimes \ket{N_B-2,2}$,
and, finally, a state where one particle 
in each component is excited 
$\ket{N_A-1,1} \otimes \ket{N_B-1,1}$.
The interactions will (partially) 
break this manifold of degenerate eigenstates.
Each of these states, once populated in the initialization step,
will induce a breathing oscillation of a characteristic frequency.
Together they represent a first order breathing manifold.
To get an insight how the respective frequencies behave
depending on the system's interactions we
briefly summarize and complement the results obtained in \cite{pyzh2018spectral}.

\begin{figure}[ht]
	\centering
	\includegraphics[width=1\columnwidth]{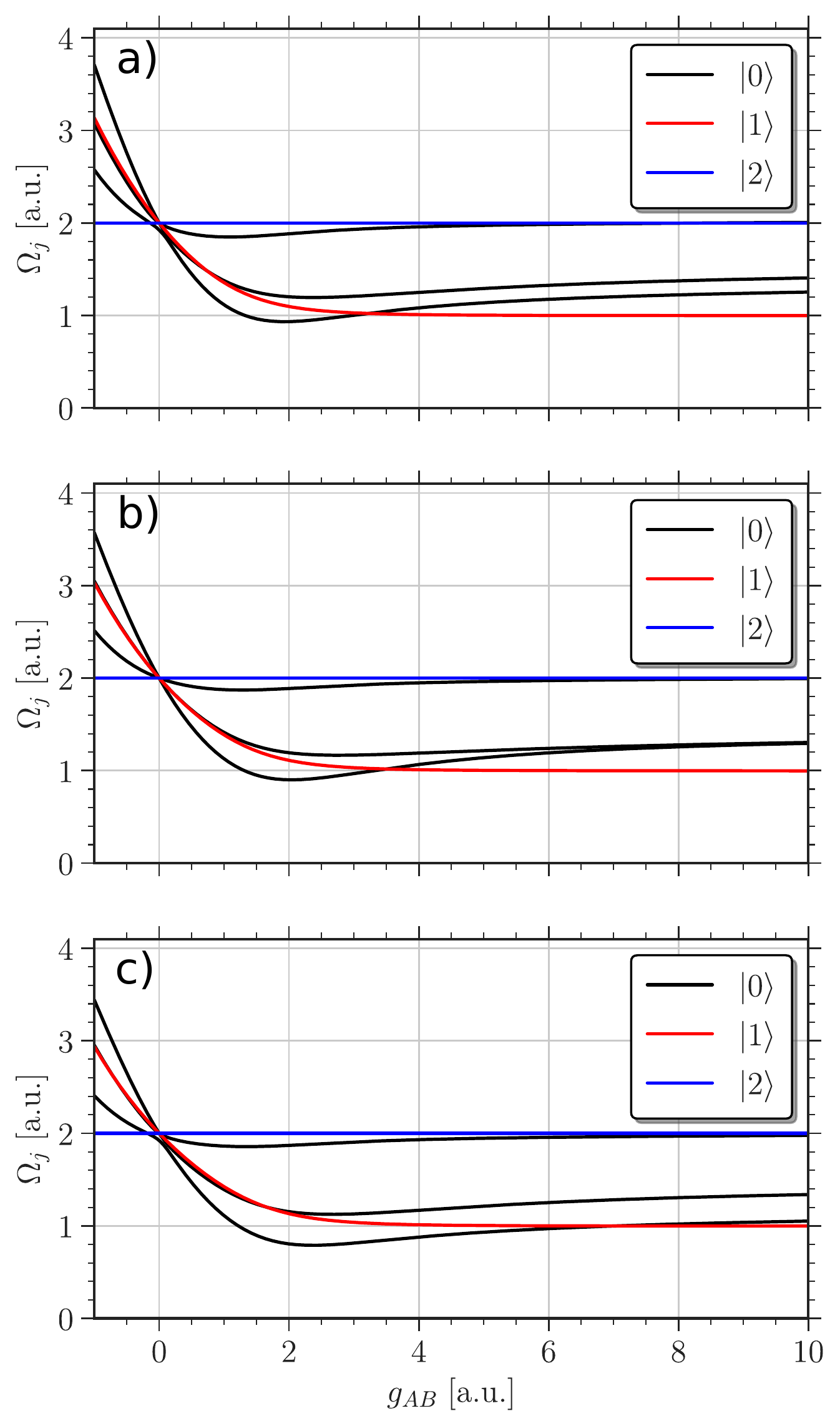}\hfil
	\caption{Breathing mode frequencies $\Omega_{j}=E_j-E_0$ 
		(w.r.t. the ground state $\ket{E_0}$)
		of a few-body bosonic mixture $N_A=2$ and $N_B=2$ 
		as a function of the inter-component coupling $g_{AB}$
		at equal trapping frequency ratio $\eta=1$, 
		intra-component interaction strength $g_A=0$ for the first component
		and a) $g_B=-0.5$, b) $g_B=0$, c) $g_B=0.5$ for the second component.
		Whether the modes are actually excited depends on the quench protocol.
		Different colors refer to the center-of-mass (CM) quantum number
		in the eigenstate $\ket{E_j}$. 
		The CM is a decoupled degree of freedom in this harmonic confinement.}
	\label{fig:bench_gs}
\end{figure}

In \cite{pyzh2018spectral}, each component consists of two particles
trapped within the same parabolic confinement $\eta=1$.
Both components experience a sudden but weak trap quench 
of the same magnitude and the system's response is studied
for different intra- and inter-component interaction strengths.
Importantly, the CM motion decouples 
while the quench operator
prevents transitions among eigenstates 
possessing different CM parity.
Four breathing mode frequencies 
have been identified and analyzed.
One of them is constant $\Omega_{CM}=2$. 
The remaining three are sensitive to interactions 
and have been labeled
$\Omega_{+}$, $\Omega_{-}$ ($\Omega_{A}$, $\Omega_{B})$ 
for species-(anti)symmetric parameter choice, 
i.e., $g_A=g_B \, (g_A \neq g_B)$, and $\Omega_{AB}$, 
all related to the relative motion of the particles.

In \cref{fig:bench_gs} the breathing frequencies 
of the above-mentioned modes are displayed
as a function of the inter-component coupling $g_{AB}$
for three different intra-component interaction values $g_B$ 
(subfigures a-c),
assuming $g_A=0$.
The curve colors encode the CM quantum number 
of the responsible eigenstates $\ket{E_j}$,
while the reference state $\ket{E_{\rm ref}}$
is the ground state $\ket{E_0}$ of even CM parity.
The blue curve is the $\Omega_{CM}$ mode 
and is indeed constant for any values
$g_A$, $g_B$ and $g_{AB}$.
The three black curves 
are the above-mentioned interaction-sensitive relative modes.
All of them have a common signature, 
namely a single minimum 
though at different values of $g_{AB}$ 
depending on the choice of intra-component interactions.
The one with the largest frequency 
is the $\Omega_{AB}$ mode.
It is quite shallow, weakly affected by
intra-component interactions and degenerates with $\Omega_{CM}$
at strong coupling $g_{AB}$.
The two lower ones are very sensitive 
to intra-component interactions and, 
depending on the presence of the species-exchange symmetry\footnote{
	corresponding to a map $x_i \mapsto y_i$ at $g_A=g_B$},
experience a bending in the vicinity of avoided crossings 
(\cref{fig:bench_gs}a and \cref{fig:bench_gs}c at $g_{AB} \approx 0$).
Finally, the red curve represents a mode caused by an eigenstate 
of odd CM parity but even global parity. 
We observe a monotonous decrease 
of the mode frequency with increasing coupling $g_{AB}$ 
until it energetically separates 
from one of the relative modes at $g_{AB}\approx1$ 
and, finally, saturates to
a value of $1.0$ at $g_{AB}\geq4$.
This mode bears strong resemblance
to one of the interaction sensitive modes
for the case of two distinguishable particles
discussed in the beginning of this section.
Its frequency is mainly affected by $g_{AB}$ and
barely by $g_{\sigma}$.
The detection of this breathing frequency requires
a numerical approach capable to incorporate the entanglement 
between the components (beyond species mean-field)
and a preparation scheme which initializes a state
having a sizable overlap with the corresponding eigenstate of odd CM parity.

\begin{figure}[h!]
	\centering
	\includegraphics[width=1\columnwidth]{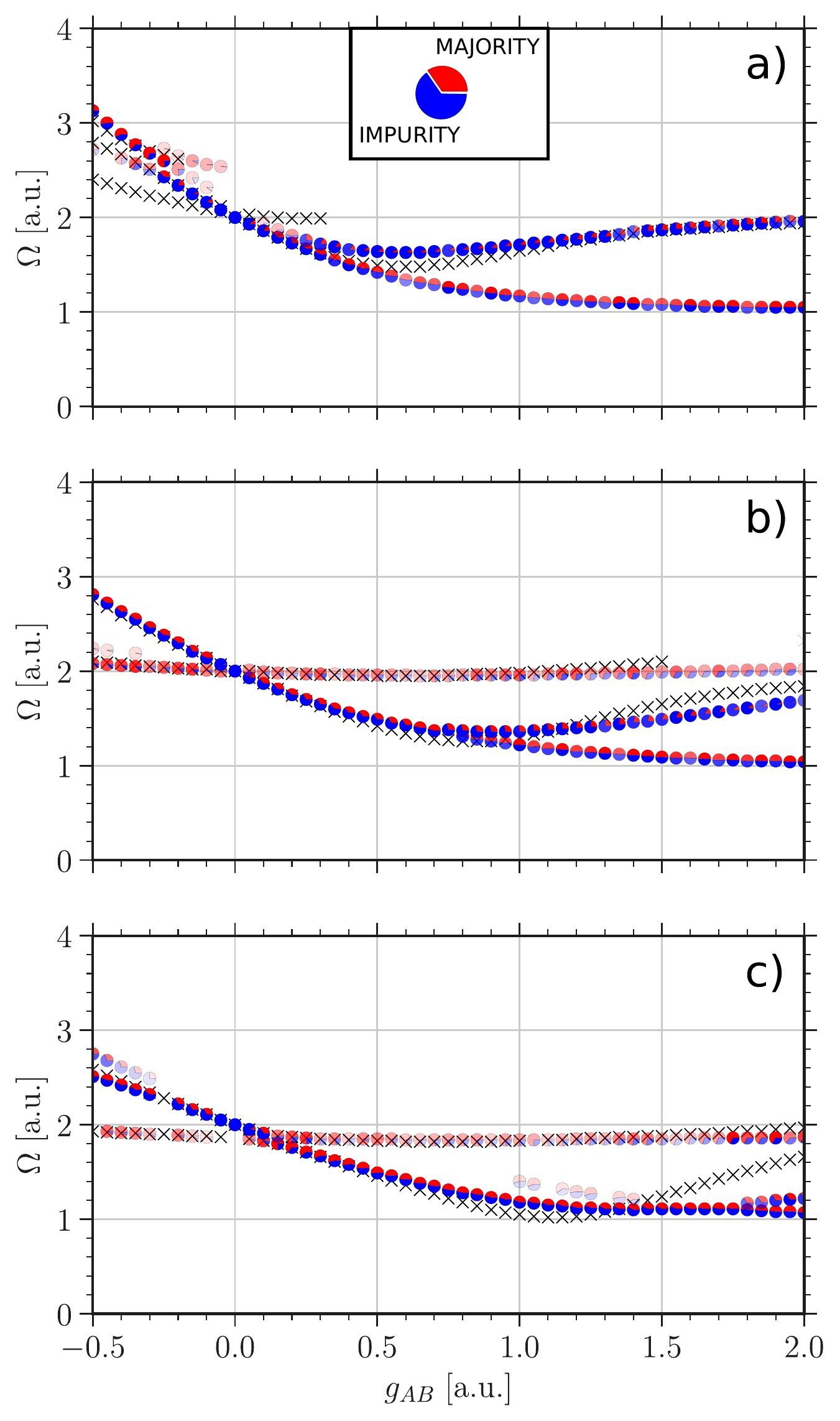}\hfil
	\caption{Frequencies $\Omega$ of breathing modes 
		excited by quenching the ground state $\ket{E_0}$
		of the Bose polaron for $N_A=5$ and $N_B=1$, 
		meaning a change in the trap ratio from $\eta=1.05$ to $\eta=1.0$,
		shown as a function of the 
		inter-component coupling $g_{AB}$ for a
		fixed majority component interaction
		a) $g_A=-0.5$, b) $g_A=0$ and c) $g_A=0.5$.
		Each frequency data point (full circle) is divided into two sectors 
		of different colors representing
		the contribution of the breathing observables
		$\braket{\sum_i \hat{x}_i^2}_t$ (red) 
		or $\braket{\hat{y}^2}_t$ (blue)			
		to the averaged power spectrum $\Sigma$ at that frequency 
		(see \cref{subsec:compressed_sensing}).
		The corresponding color intensity indicates the relative strength w.r.t. 
		the maximum amplitude $\Sigma_{\rm{max}}$
		in the averaged power spectrum for fixed $g_{AB}$ and only
		modes with contribution above $10\%$ of $\Sigma_{\rm{max}}$ 
		are presented. 
		Crosses stand for frequencies of modes excited 
		within the SMF approximation.}
	\label{fig:bose_polaron_5_1_eta_1}
\end{figure}

\subsection{Bose polaron}
\label{subsec:polaron_breathing}

Now we turn our attention to a single impurity $N_B=1$ 
in a few-body majority environment 
having $N_A=5$ or $N_A=10$ particles.
In contrast to \cite{pyzh2018spectral},
here, we relax only the $B$ component, while the $A$ component
is affected indirectly via the inter-component coupling $g_{AB}$.
At $\eta=1$ the CM motion still decouples from the relative motions.
In particular, the quench operator may mediate
between eigenstates of different CM parity
inducing eventually a special breathing mode 
caused by population of an eigenstate with odd CM parity.
Furthermore, there is only one particle in the $B$ component.
Thus, a double excitation state $\ket{N_A} \otimes \ket{N_B-2,2}$ 
(see \cref{subsec:bose:mixture})
does not exist and 
we expect that one of the relative modes 
(black curves in \cref{fig:bench_gs}), 
whose frequency is notably affected by interactions, 
will not be present.

\begin{figure}[h!]
	\centering
	\includegraphics[width=1\columnwidth]{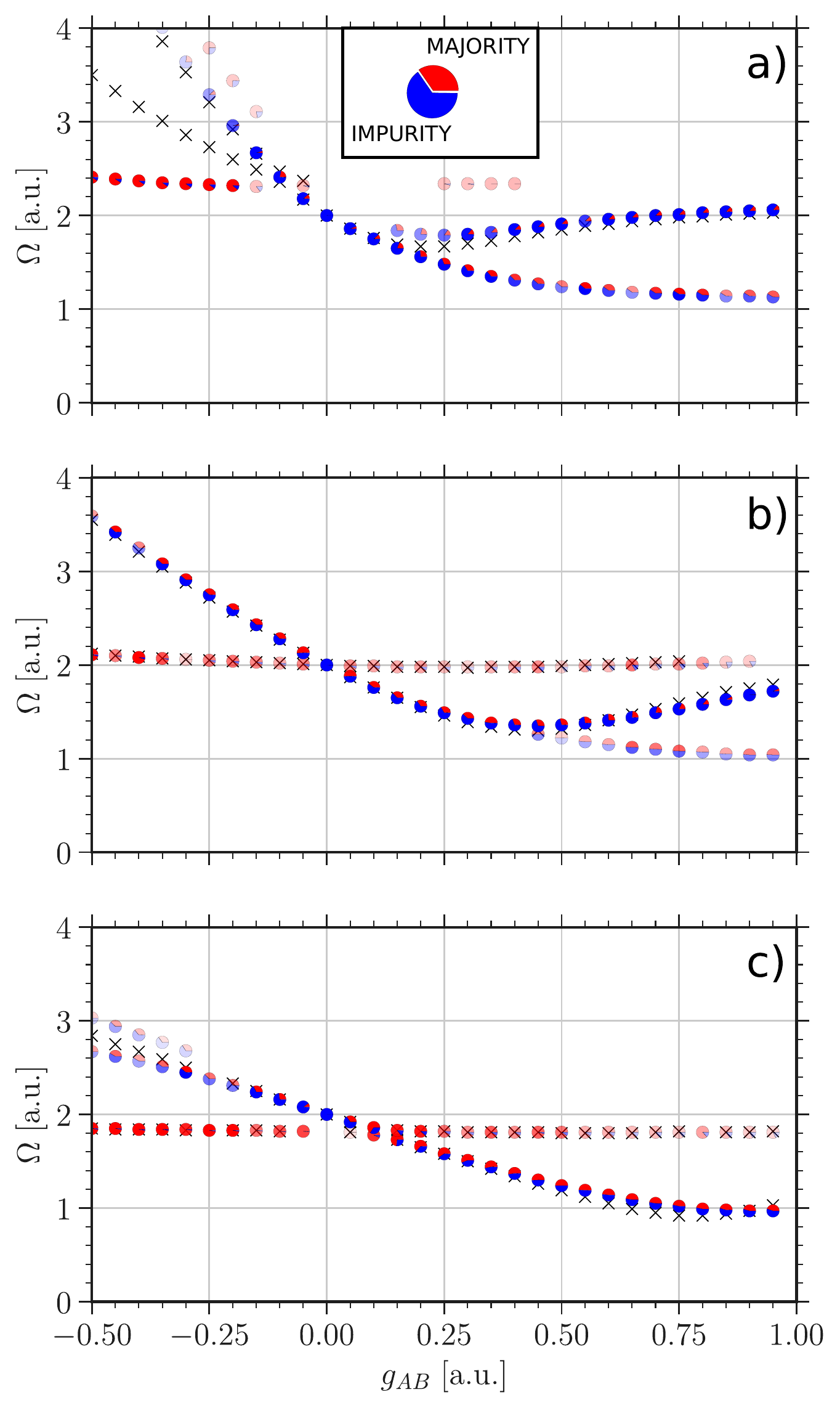}\hfil
	\caption{Same as in \cref{fig:bose_polaron_5_1_eta_1} 
		but for $N_A=10$.}
	\label{fig:bose_polaron_10_1_eta_1}	
\end{figure}

In \cref{fig:bose_polaron_5_1_eta_1,fig:bose_polaron_10_1_eta_1}
we show the excitation spectrum of the breathing dynamics initialized
by quenching the equilibrated system at trap ratio $\eta=1.05$ to $\eta=1$,
i.e., partially releasing the trap of the impurity. 
The majority component consists
of $N_A=5$ (\cref{fig:bose_polaron_5_1_eta_1})
or $N_A=10$ (\cref{fig:bose_polaron_10_1_eta_1}) particles subject to
several different majority component interactions $g_A$ (subfigures a-c).
Only frequencies of modes whose contribution is above $10\%$ of the maximum amplitude
$\Sigma_{\rm max}$ in the averaged power spectrum are shown.
Additionally, each frequency data point (full circle) is represented as a pie chart
of two different colors and encodes the contribution
of the breathing observables to the averaged power spectrum
(see \cref{subsec:compressed_sensing}):
blue color for the impurity $\hat{y}^2$ and
red color for the  majority component $\sum_i \hat{x_i}^2$.
The decomposition into colors tells us whether the respective mode is
a single species mode or whether 
it is of a mixed character and to what extent.
Furthermore, the color intensity indicates the
participation of the respective mode in the breathing dynamics
as compared to the most relevant mode at fixed $g_{AB}$ 
(a more intense color indicates a stronger contribution).
Finally, the crosses represent frequencies of modes
excited by the same procedure but numerically ignoring the entanglement
in the initial state and the subsequent dynamics (SMF approximation).

Let us first focus on \cref{fig:bose_polaron_5_1_eta_1}b, 
the case of a non-interacting majority species ($g_A=0$).
At $g_{AB}=0$ only the impurity is excited (blue circle)
performing a breathing motion with frequency $\Omega=2$ as expected.
As one increases the coupling strength ($g_{AB}>0$), a
second mode of decreasing frequency emerges resulting
in a beating.
This mode has the largest contribution to the ongoing dynamics
and is more prominent in the impurity motion.
The frequency of the other mode experiences 
only a slight variation
$\Omega \approx 2$
and is represented to a larger extent in the majority component.
It matches the behavior of the relative mode
least sensitive to interactions
encountered in \cref{subsec:bose:mixture}
(cf.\ black curves in \cref{fig:bench_gs}).

Around $g_{AB}\approx 0.75$ the lower frequency splits into two branches 
of comparable significance
resulting altogether in three modes.
One of the emerging branches possesses a continuously
decreasing frequency with increasing $g_{AB}$ approaching the value $\Omega=1$.
It matches the description of the inter-component mode 
which is caused by an eigenstate of odd CM parity and was
emphasized in \cref{subsec:bose:mixture}
(cf.\ red curve in \cref{fig:bench_gs}).
Interestingly, it is equally represented 
in both subsystems despite
the particle number imbalance and quench asymmetry.
The other frequency branch bends and starts recovering towards $\Omega=2$
with increasing $g_{AB}$ while gradually becoming
a pure signature of the impurity motion only 
(the blue slice dominates at $g_{AB}=2$).
It reminds of the two convex mode frequencies particularly sensitive to interactions
mentioned in \cref{subsec:bose:mixture}
(cf.\ the black curves in \cref{fig:bench_gs}).

At weak negative coupling ($g_{AB}<0$) 
one observes also a beating behavior,
although here both frequencies are increasing with decreasing $g_{AB}$.
The dominant frequency in the power spectrum
is more sensitive to the coupling variation 
and dominates the impurity motion,
while the second is barely affected and primarily 
represented in the majority component.
Below a certain threshold ($g_{AB}<-0.4$) one observes 
low-amplitude traces of a third frequency.

We discover several major alternations in the excitation spectrum
when the majority component becomes interacting.
For weakly attractive ($g_A=-0.5$) majority particles
in \cref{fig:bose_polaron_5_1_eta_1}a
the coupling-insensitive mode frequency is seemingly absent.
At positive increasing $g_{AB}$ 
the bifurcation of the continuously decaying frequency
takes place already for
a very weak coupling strength ($g_{AB} \approx 0.1$).
The character of the excited modes is mostly the same as for $g_{A}=0$.
At negative decreasing $g_{AB}$ we observe an emerging multi-frequency breathing
composed of three modes.
The lower-frequency mode looses amplitude in favor
of higher-frequency modes ($-0.4<g_{AB}<-0.1$).
Then it turns into a single-frequency breathing ($g_{AB}<-0.4$)
affecting both components in a similar way.

For weakly repulsive ($g_A=0.5$) majority particles
in \cref{fig:bose_polaron_5_1_eta_1}c
the coupling-insensitive mode frequency 
is still present though energetically shifted
downwards to $\Omega=1.9$.
The respective mode is weakly represented in the overall dynamics
and affects mainly the majority component.
The point of bifurcation in the lower frequency branch is located 
at a stronger coupling $g_{AB}\approx1.8$.
At negative $g_{AB}$ below a certain threshold ($g_{AB}<-0.3$)
a third mode is excited.
In contrast to $g_A=0$, this additional mode is 
rather manifested in the impurity breathing and has a larger frequency.
In summary, the majority component interaction 
$g_A$ determines the coupling value 
at which the bifurcation takes place
as well as the offset of the coupling-insensitive frequency 
and whether it can be addressed by the current quench protocol.

Next, we double the number of majority component atoms
to get an idea of how it affects the excitation spectrum.
In the following, we compare the corresponding subfigures
of \cref{fig:bose_polaron_5_1_eta_1,fig:bose_polaron_10_1_eta_1}.
We note that at $N_A=10$ the interval of the considered couplings 
is $g_{AB} \in [-0.5,1.0]$
as the convergence is more challenging to achieve 
beyond $g_{AB}>1$.
At $g_A=0$ (\cref{fig:bose_polaron_10_1_eta_1}b)
the bifurcation point is located at a smaller value of $g_{AB}$
compared to the $N_A=5$ case (see \cref{fig:bose_polaron_5_1_eta_1}b).
After the bifurcation, 
we observe a decreasing relevance of the
lower frequency mode 
in favor of the upper branch.
We anticipate that its amplitude will decay further 
for an even larger number of majority atoms.
The coupling-insensitive mode frequency is barely affected 
by the particle number imbalance.

At $g_A=-0.5$ (\cref{fig:bose_polaron_10_1_eta_1}a) 
and positive $g_{AB}$ 
the minimum of the convex frequency mode is
shifted to larger frequencies 
and smaller values of the coupling $g_{AB}$.
The frequency value recovers back to $\Omega=2$
more quickly already at $g_{AB}\approx 0.75$.
The coupling-insensitive mode becomes visible at negative $g_{AB}$
and even dominates the breathing dynamics,
although the amplitude
decays considerably towards $g_{AB}=0$ 
and there are only minor traces left
at positive coupling ($0.25<g_{AB}<0.4$).
It is certainly present in the breathing dynamics at positive $g_{AB}$,
but the contribution is not significant enough 
to overcome the set threshold.
The respective frequency is shifted to $\Omega \approx 2.4$. 
At $g_A=0.5$ (\cref{fig:bose_polaron_10_1_eta_1}c)
the frequency of the coupling-insensitive mode experiences 
a slight shift downwards.
At negative $g_{AB}$ it gains amplitude with decreasing $g_{AB}$
until it becomes a dominant mode below $g_{AB}\leq-0.25$.
The bifurcation point at positive $g_{AB}$ is 
unfortunately not visible within 
the covered $g_{AB}$ interval.

Lastly, we want to address the importance of entanglement in
our Bose polaron setup. 
To this end we neglect it both in the initial state 
and in the subsequent dynamics 
(crosses in \cref{fig:bose_polaron_5_1_eta_1,fig:bose_polaron_10_1_eta_1}). 
The first striking observation is that in this case at most
{\it two} frequencies can be extracted.
The mode we are missing from the exact simulations 
is the one whose frequency 
is a monotonically decreasing function of $g_{AB}$.
Regarding the persistent modes, the one with a
coupling-insensitive frequency is overall well-captured by
the species mean-field ansatz, although
it tends to overestimate the frequency for large positive values of $g_{AB}$.
The other one, whose frequency is a convex function of $g_{AB}$,
in general fails to match the corresponding exact frequency:
either mispredicting the location (\cref{fig:bose_polaron_5_1_eta_1}c)
or the exact value (\cref{fig:bose_polaron_5_1_eta_1}a) of the minimum.
Even if both the location and the value of the minimum 
are well-matched (\cref{fig:bose_polaron_5_1_eta_1}b)
there is an increasing discrepancy for strong positive $g_{AB}$.
For a larger particle number (\cref{fig:bose_polaron_10_1_eta_1})
the consistency between approximated and exact frequencies
is much better, though also here one mode is missing.

To summarize, we are able to excite 
up to three breathing modes 
in the Bose polaron setup by quenching only the impurity.
First, there is a coupling-insensitive mode frequency 
whose value can be manipulated
by the particle number imbalance 
or the majority component interaction.
Second, there is a monotonically decreasing frequency
converging towards $\Omega=1.0$ with increasing $g_{AB}$.
It cannot be described by the species mean-field ansatz
and its contribution to the ongoing dynamics decreases
with larger particle number imbalance.
Third, at a sufficiently strong coupling 
an additional mode enters the dynamics.
The mode frequency is a convex function of $g_{AB}$ 
with a minimum being sensitive to $g_A$ and $N_A$.

\subsection{Impact of the trap}
\label{subsec:inhomogeneity}

Let us now focus on the impact of the external trap, more specifically
we consider a situation 
where the length scale of the impurity 
$l_B=\sqrt{1/\eta}$, set
by the parabolic trap, is either broader  
($l_B=1.4$)
or narrower ($l_B=0.5$) in the post-quench system. 
The quench strength is still $5\%$ of the original trap parameter.

\begin{figure}[ht]
	\centering
	\includegraphics[width=0.5\textwidth]{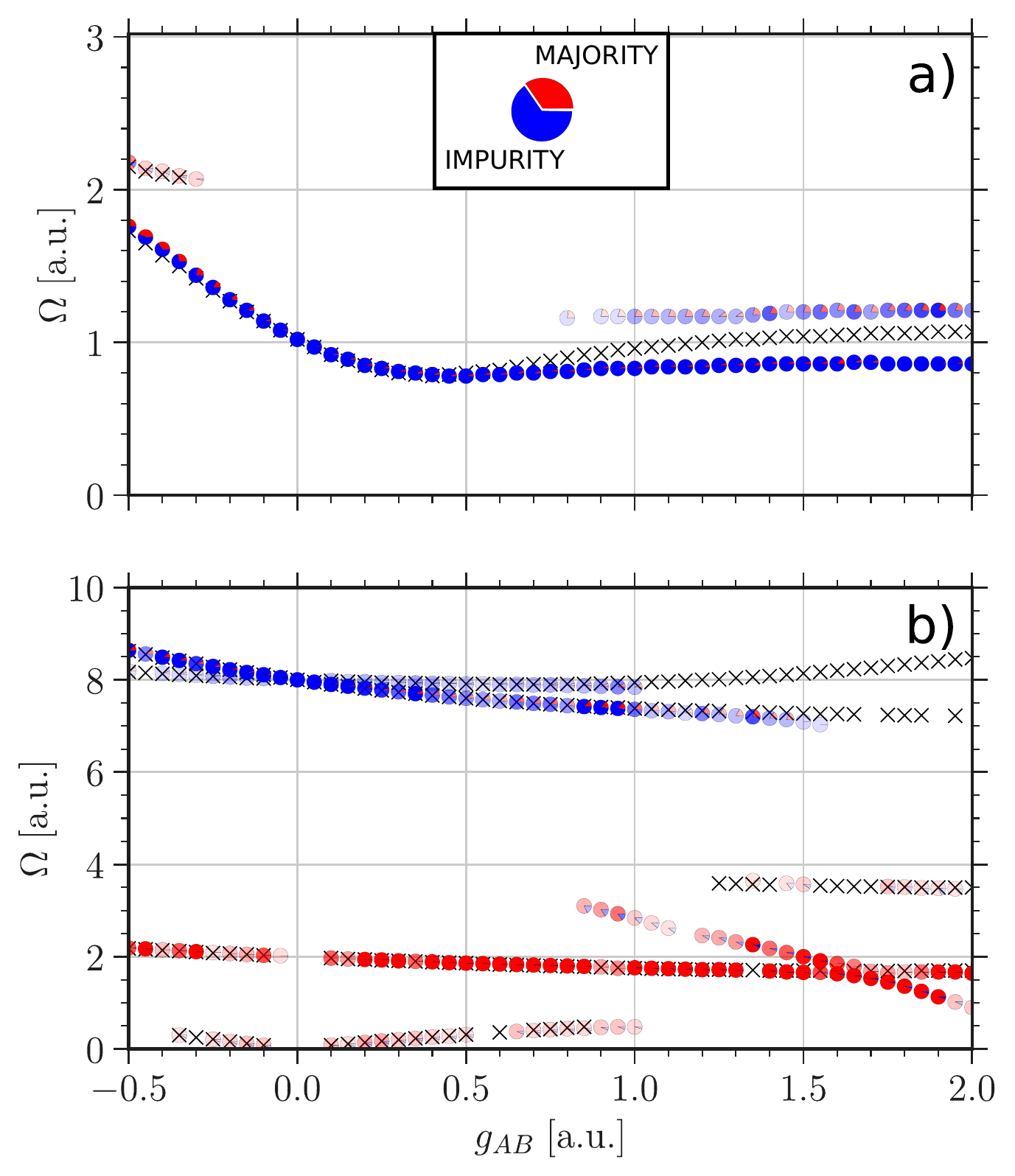}\hfil
	\caption{Frequencies $\Omega$ of breathing modes 			
		for $N_A=5$ and $N_B=1$ 
		as a function of the 
		inter-component coupling $g_{AB}$ 
		at fixed majority component interaction $g_A=0$.
		The trap ratio is quenched in (a) from $\eta=0.536$ to $\eta=0.51$ and
		in (b) from $\eta=4.2$ to $\eta=4$.
		Color coding according to \cref{fig:bose_polaron_5_1_eta_1}.}
	\label{fig:bose_polaron_5_1_gA_0}	
\end{figure}

\begin{figure}[ht]
	\centering
	\includegraphics[width=0.5\textwidth]{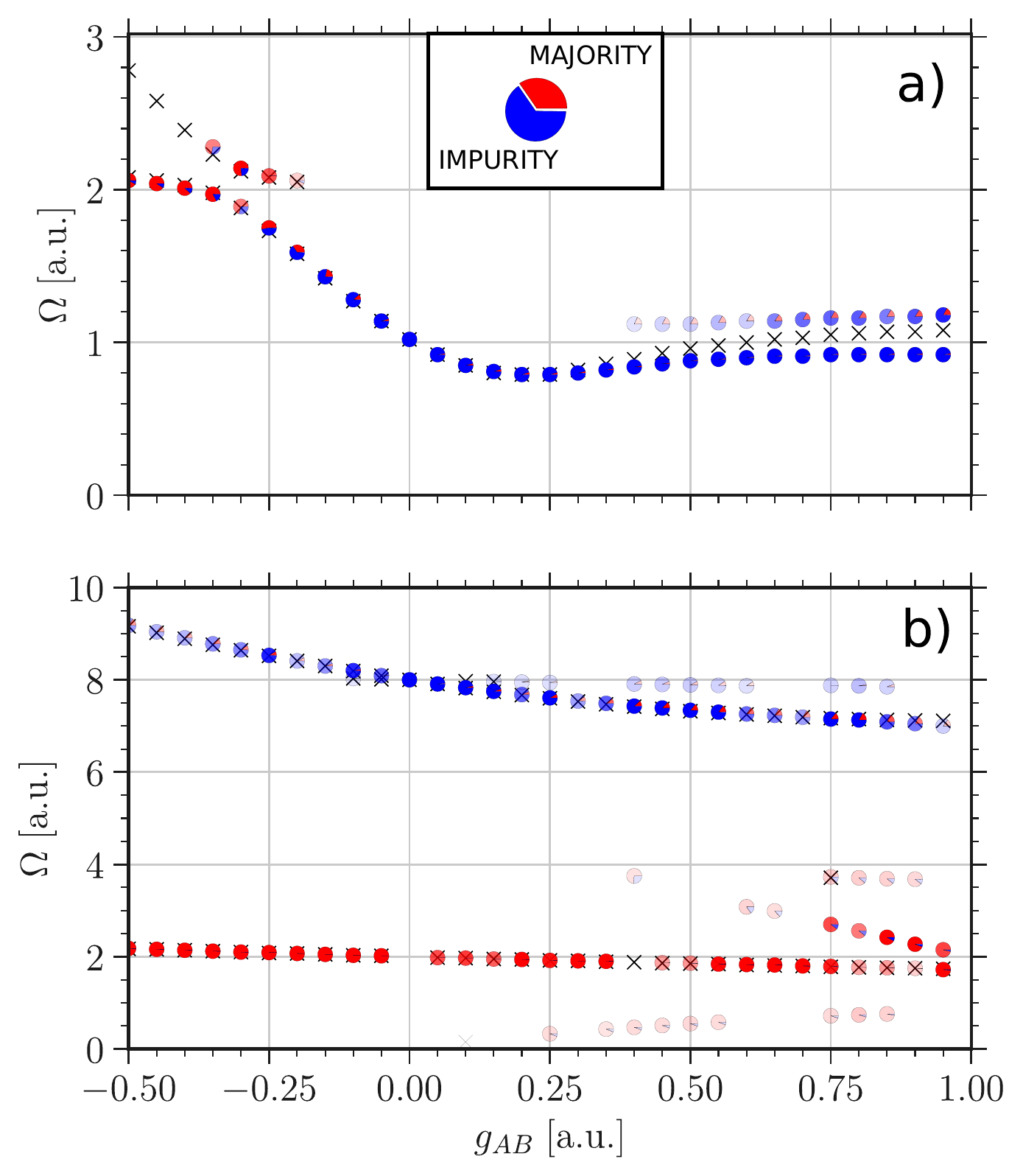}\hfil
	\caption{Same as in \cref{fig:bose_polaron_5_1_gA_0} 
		but for $N_A=10$.}
	\label{fig:bose_polaron_10_1_gA_0}	
\end{figure}

We start with the case of a 'broad' impurity $N_B=1$.
In \cref{fig:bose_polaron_5_1_gA_0}a 
and \cref{fig:bose_polaron_10_1_gA_0}a 
we show the breathing spectrum
for $N_A=5$ and $N_A=10$ majority atoms, respectively.
To gain an intuitive picture we set $g_{A}=0$.
For the decoupled case $g_{AB}=0$, the lowest frequency mode
is caused by the eigenstate $\ket{N_A} \otimes \ket{0,0,1}$, 
corresponding to a standard breathing
of the impurity at frequency $\Omega=1.02$.
It is the only mode excited.
Once coupled ($g_{AB}\neq0$), several other eigenstates may become populated
leading to additional breathing modes.
The states in question can be continiously traced back to 
the low-energy eigenstates of a decoupled impurity.
First, we have the state $\ket{N_A-1,1} \otimes \ket{0,1}$ 
corresponding to an excitation of a sloshing mode in each component
at frequency $\Omega=1.51$.
Then follows a quasi-degenerate manifold of three modes: 
two majority component modes at the same frequency $\Omega=2$ 
caused by $\ket{N_A-1,0,1} \otimes \ket{1}$ and $\ket{N_A-2,2} \otimes \ket{1}$,
and a second order breathing of the impurity at frequency $\Omega=2.04$
mediated by $\ket{N_A} \otimes \ket{0,0,0,0,1}$.
Any higher frequency modes are unlikely to be involved.

At weak coupling $g_{AB}$ there 
is only one relevant mode excited.
It originates from the state $\ket{N_A} \otimes \ket{0,0,1}$ and is
barely detectable in the majority component breathing.
The corresponding frequency is a convex function of $g_{AB}$
with a minimum at $g_{AB}\approx0.5$ 
for $N_A=5$ (\cref{fig:bose_polaron_5_1_gA_0}a)
and at $g_{AB}\approx0.25$ 
for $N_A=10$ (\cref{fig:bose_polaron_10_1_gA_0}a).
At strong positive $g_{AB}$ a beating behavior emerges.
The amplitude of the additional mode 
increases gradually with increasing $g_{AB}$ while
the corresponding frequency $\Omega \approx 1.2$ is only weakly 
affected by the inter-component coupling or the particle number.
At negative moderate $g_{AB}$ there is also a beating.
The major amplitude mode, originally being an impurity mode (blue),
evolves gradually into the majority component mode (red).
The other mode appears just below $g_{AB}<-0.25$ and
affects primarily the majority species.

The SMF fits well the lowest frequency at negative $g_{AB}$
and at weak positive $g_{AB}$ until the minimum is reached.
Afterwards, it overestimates the frequency having
larger deviations at stronger positive $g_{AB}$.
Again, we witness that SMF is incapable
to identify an emerging mode at positive $g_{AB}$,
though at negative $g_{AB}$ 
it does register the beating behavior.
It implies that the additional modes entering the dynamics 
at positive and negative $g_{AB}$
are of a different character.
Based on the insights gained in the previous section,
namely the presence of a breathing mode
which is inaccessible to the SMF treatment,
we conjecture that the higher frequency mode observed at positive $g_{AB}$
stems from the doubly excited sloshing mode $\ket{N_A-1,1} \otimes \ket{0,1}$
at $g_{AB}=0$.
Its contribution grows as the entanglement becomes stronger.

Next, let us focus on the case of a 'narrow' impurity $N_B=1$.
The corresponding breathing spectrum is depicted 
in \cref{fig:bose_polaron_5_1_gA_0}b for $N_A=5$
and in \cref{fig:bose_polaron_10_1_gA_0}b for $N_A=10$
majority atoms.
At $g_{AB}=0$ we excite only the standard breathing mode of the impurity
at frequency $\Omega=8$ caused by the eigenstate $\ket{N_A} \otimes \ket{0,0,1}$.
Considering the amount of available even parity eigenstates
with energies up to eight quanta
($40$ for $N_A=5$ and $45$ for $N_A=10$),
one might naively think that many modes would be excited at finite $g_{AB}$. 
This is not the case as we count only up to six frequencies.
They are well separated from each other and 
admit a convenient classification: 
impurity modes (blue) with $\Omega>6$
and majority modes (red) with $\Omega<4$.

The impurity features a beating composed of two modes at weak coupling.
The gap between the corresponding frequencies grows with increasing $g_{AB}$.
The one of smaller amplitude 
vanishes around $g_{AB}\approx1.0$.
The contribution of the other mode fades away quickly afterwards
until it also disappears.
At strong $g_{AB}$
the impurity motion assimilates the majority component breathing.
Both modes are reproducible by SMF ansatz,
though SMF overestimates their contribution 
to the overall dynamics at strong $g_{AB}$.

Regarding the majority modes there is one
with a nearly constant frequency ($\Omega \approx 2$) 
entering the dynamics already at weak coupling
and making a large contribution 
to the majority motion across all coupling values.
At weak $g_{AB}$ it is accompanied by an oscillation of a smaller frequency.
As the ground state is non-degenerate at $g_{AB}=0$,
this frequency corresponds to 
the gap between the two blue-colored frequencies.
It also consistently disappears beyond $g_{AB}>1$ 
along with the impurity modes.
The latter are actually replaced
by modes of lower frequency.
One of them is of particular interest.
It appears at $g_{AB}\approx1$ for $N_A=5$
and at $g_{AB}\approx0.5$ for $N_A=10$
gaining weight with increasing $g_{AB}$.
The corresponding frequency is a linearly decreasing function of $g_{AB}$.
It can be interpolated to frequency $\Omega=5$ at $g_{AB}=0$,
matching the energy gap between the ground state $\ket{N_A} \otimes \ket{1}$ and
the inter-component sloshing mode eigenstate $\ket{N_A-1,1} \otimes \ket{0,1}$.
The entanglement is once again indispensable to account 
for the respective breathing mode.

\FloatBarrier

\subsection{Breathing of the first excited state}
\label{subsec:1excbreathing}

The Hamiltonian \cref{eq:hamiltonian}
has global reflection symmetry.
The eigenstates are therefore separable into two classes
of even and odd global parity.
The quench operator does not violate that symmetry.
Accordingly, an even parity initial state can be
expanded within the subspace of even eigenstates.
The odd parity space of the Hamiltonian
has its own 'ground state', 
meaning the lowest energy eigenstate of that subspace.
If initialized in such a state,
how will each species respond following our quench procedure? 
Will it be a few-mode breathing
within each component,
as for the even parity ground state, 
or a more complex motion involving many modes?
If only a few modes participate, how different are the
respective frequencies as compared to the even parity ground state?

\begin{figure}[h!]
	\centering
	\includegraphics[width=1\columnwidth]{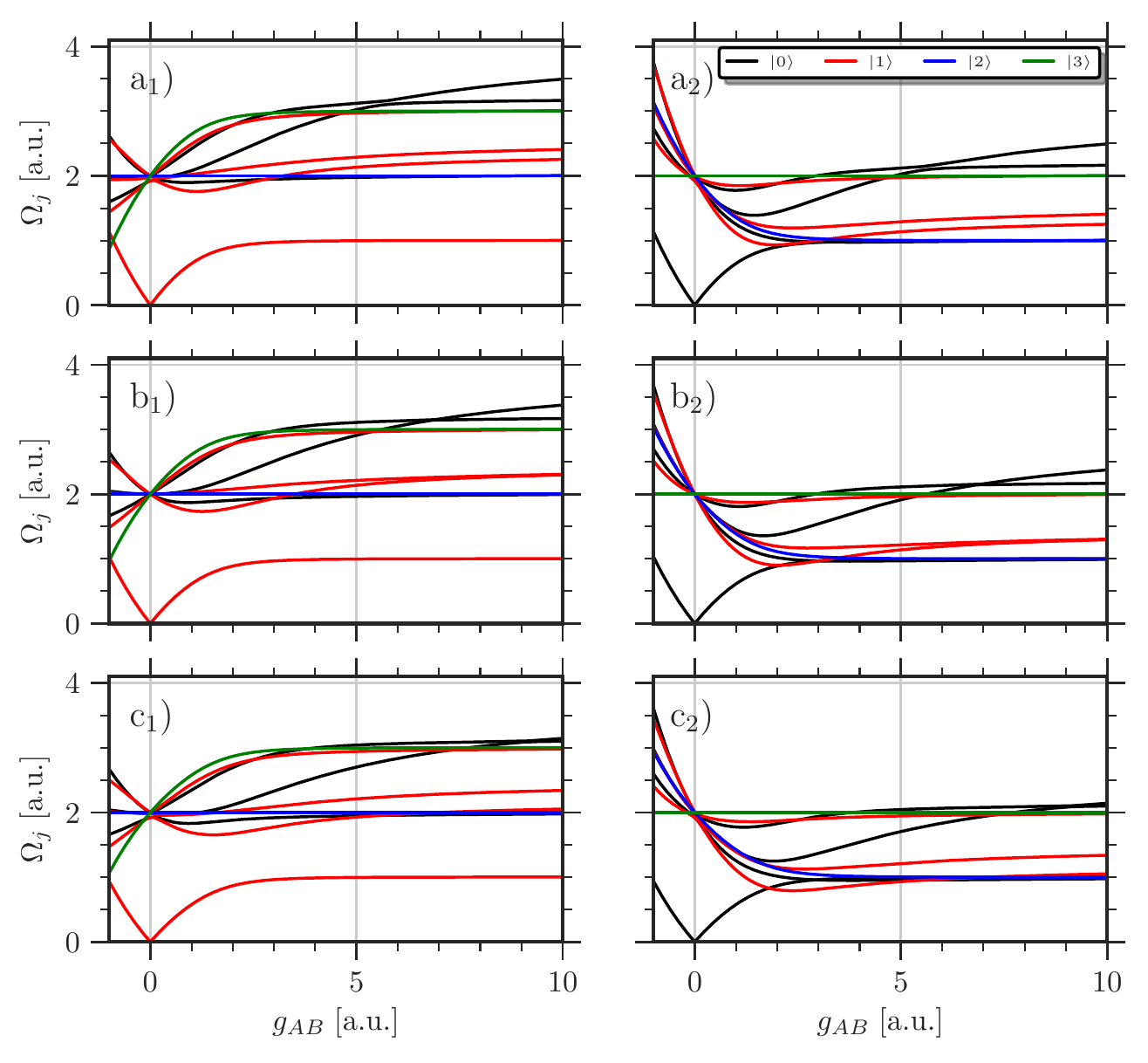}\hfil
	\caption{Energy gaps $\Omega_{j}=|E_j-E_{\rm{ref}}$|
		w.r.t. lowest energy reference eigenstates $\ket{E_{\rm{ref}}}$ 
		of odd global parity and even (first column) 
		or odd (second column) CM parity
		in a few-body bosonic mixture $N_A=N_B=2$.
		The gaps are functions of the inter-component coupling $g_{AB}$
		at equal trapping frequency ratio $\eta=1$, 
		intra-component interaction strength $g_A=0$ for the first component
		and a) $g_B=-0.5$, b) $g_B=0$, c) $g_B=0.5$ for the second component.
		Whether the modes are actually excited depends on the quench protocol.
		Different colors refer to the center-of-mass (CM) quantum number
		in the eigenstate $\ket{E_j}$. 
		The CM is a decoupled degree of freedom in this harmonic confinement.}
	\label{fig:bench_exc}	
\end{figure}

To address the above questions,
we consider again the example of a particle-balanced
few-body Bose mixture (see \cref{subsec:bose:mixture}) at $\eta=1$.
We notice that
in the non-interacting regime
the odd parity 'ground state' is two-fold degenerate, 
composed of states
where a single particle of either component is excited 
by one energy quantum:
$\ket{N_A-1,1} \otimes \ket{N_B}$ and $\ket{N_A} \otimes \ket{N_B-1,1}$
w.r.t. harmonic oscillator basis.
Once the degeneracy is lifted at finite coupling
the perturbed eigenstates can be distinguished by the CM quantum number.
As both of them are likely to be populated after the quench,
we account for both as reference states when evaluating the energy gaps 
to neighboring eigenstates.
The frequencies of breathing oscillations following a quench of the impurity trap
will be contained within the set of these energy gaps.
At zero interactions,
the energetically closest potentially accessible neighboring eigenstates are
the three quanta excitations.
There are eight of them in total.
Half of them are excitations within a single component:
$\ket{N_A-1,0,0,1}$, $\ket{N_A-2,1,1}$ 
and the same for $B$ species.
The other four distribute the three available quanta
over both components:
$\ket{N_A-1,1} \otimes \ket{N_B-2,2}$, $\ket{N_A-1,0,1} \otimes \ket{N_B-1,1}$
and the other way around ($B \leftrightarrow A$).

In \cref{fig:bench_exc} we show the energy gaps
between a reference state 
of even or odd CM parity (first or second column respectively)
and energetically closest eigenstates (CM quantum number indicated by color)
as a function of the inter-component coupling $g_{AB}$
for three different intra-component interaction regimes (rows).

Let us begin with the reference state of even CM parity (first column).
First, there is a single constant frequency mode $\Omega=2$ (blue) 
for any interaction values.
Second, there is a single frequency being
a monotonically increasing function of $g_{AB}$ (green)
and saturating to $\Omega=3$ at strong positive $g_{AB}$.
Third, among the three black curves there is one
very weakly dependent on the interactions and it recovers to $\Omega=2$
at strong positive $g_{AB}$, whereas the other two
are highly susceptible to interactions and reach values
beyond $\Omega=3$.
Finally, the lower red curve represents
the other reference state of odd CM parity.
Regarding the frequencies of the remaining three red modes, 
one of them behaves similar to the green mode,
while the other two are highly sensitive to interactions.
Some of the crossings seen at $g_{B}=0$ (subfigure b)
among the black and red curves
become avoided at finite $g_{B}$ (subfigures a and c)
caused by broken species exchange symmetry.
Overall, most frequencies reach values above $\Omega=2$
and there is nothing common to the breathing spectrum 
of the even CM-parity ground state
(see \cref{fig:bench_gs}) except the constant frequency mode.

Focusing now on the reference state of odd CM-parity (second column),
we notice that all five frequencies encountered in \cref{fig:bench_gs}
have here a corresponding match.
The reason is that the reference state is a simple CM excitation, being
a constant energy shift independent of the interaction strength.
Correspondingly, the even parity ground state and the eigenstates
responsible for the first order breathing modes 
discussed in \cref{subsec:bose:mixture} 
are just spectrally shifted by a common constant.
Thus, the corresponding energy gaps remain intact.
The four black curves are the additional new modes.
The lowest one corresponds to the even CM-parity reference state 
of odd global parity.
There is one with a monotonically decreasing frequency and two
of them are convex functions of $g_{AB}$ very sensitive to interactions
akin to the red mode frequencies.

\begin{figure}[h!]
	\centering
	\includegraphics[width=0.5\textwidth]{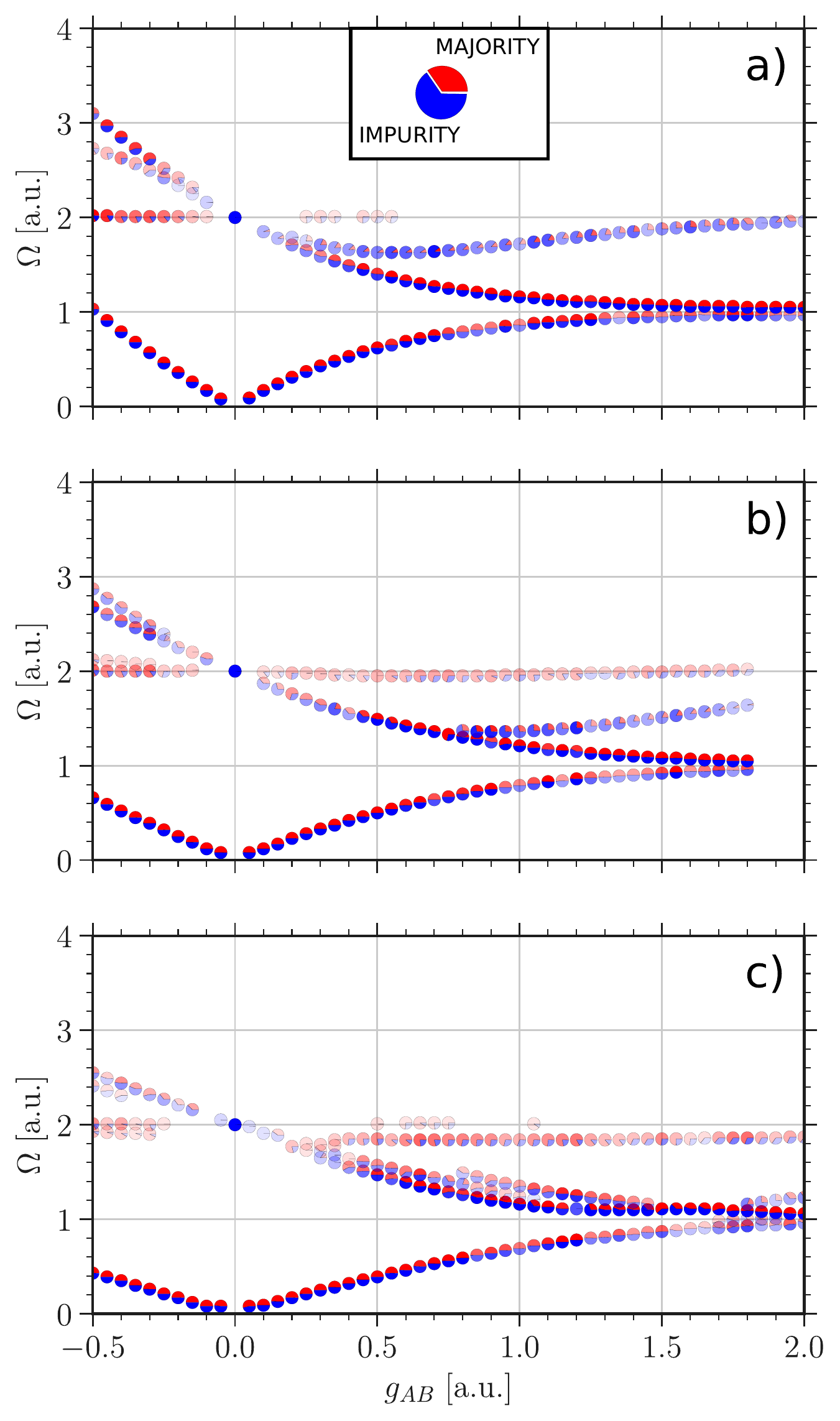}\hfil
	\caption{Frequencies $\Omega$ of breathing modes 
		excited by quenching the odd parity 'ground state' $\ket{E_1}$
		of a Bose polaron $N_A=5$ and $N_B=1$ 
		from a trap ratio $\eta=1.05$ to $\eta=1.0$
		as a function of the 
		inter-component coupling $g_{AB}$ 
		for a fixed majority component interaction
		a) $g_A=-0.5$, b) $g_A=0$ and c) $g_A=0.5$.
		Color coding as in \cref{fig:bose_polaron_5_1_eta_1}.}
	\label{fig:exc_5_1_eta_1}	
\end{figure}

In the Bose polaron setup 
there are two less 'three-quanta' states, 
since the two-particle excitations 
of the impurity are obviously excluded.
However, there is also one more state, 
namely a three-particle excitation $\ket{N_B-3,3}$ 
in the majority component.
Thus, there are seven addressable low-energy eigenstates in total.
Now, we initialize an odd parity 'ground state' 
for the subsequent breathing dynamics
and extract the frequencies 
of participating modes shown in \cref{fig:exc_5_1_eta_1}.
We immediately recognize 
the frequency pattern from \cref{fig:bose_polaron_5_1_eta_1}.
In particular, we evidence a coupling-insensitive frequency 
at $\Omega \approx 2$, 
a monotonically decaying one converging to $\Omega=1$
and one which is a convex function of $g_{AB}$.
There are however several differences.
Importantly, both odd parity 'ground states' 
are participating in the dynamics,
as indicated by the lowest frequency curve.
The excitation pattern implies that the odd-CM 'ground state'
has a larger contribution than the one of even-CM.
Second, at weak positive $g_{AB}$ 
the two odd parity reference states
are dominating the dynamics. The energy gap between them
grows with increasing $g_{AB}$ 
and approaches its limiting value $\Omega=1$.
Meanwhile, there is
a gradual transfer of population to the eigenstate responsible for
the breathing frequency which is monotonically decaying function of $g_{AB}$.
Finally, at $g_{A}=0.5$
and positive intermediate $g_{AB}\approx1$
as well as at negative $g_{AB}$
we identify some minor traces of additional modes 
absent for the even parity ground state.

\section{Summary and Conclusions} 
\label{sec:conclusions}

The breathing dynamics
of a few-body Bose polaron 
in a one-dimensional species-selective parabolic confinement 
has been investigated in this work
by means of the  Multi-Layer Multi-Configuration Time-
Dependent Hartree Method for Bosons.
The dynamics has been triggered by a weak trap quench of the impurity for
different inter-component couplings $g_{AB}$ ranging
from weak attractive to intermediate repulsive values.
The majority motion was affected indirectly 
via the majority-impurity interaction $g_{AB}$.
We extracted the frequencies of excited modes
from the breathing observables by using a
Compressed Sensing algorithm.
From this we constructed an averaged power spectrum
and classified the modes according to their overall contribution
to the dynamics.
We also determined whether a mode is of majority or impurity type 
judged by the relative strength of respective observables 
in the averaged power spectrum.
To highlight the importance of entanglement in our setup
we performed the same quench procedure 
for a species-mean field ansatz, 
which assumes that a wave-function
can be written as a single product state 
of combined majority coordinates and the impurity coordinate.

Different regimes of system parameters have been addressed
The majority component was non-interacting ($g_{A}=0$), 
weakly attractive ($g_{A}=-0.5$) or weakly repulsive ($g_{A}=0.5$) 
consisting either of $N_A=5$ or $N_A=10$ particles
and different ratios of species trapping frequencies
including equal localization length ($\eta=1$), 
a 'broad' impurity ($\eta=0.51$) and a 'narrow' impurity ($\eta=4.0$)
have been taken into account.
Finally, we studied the impact of parity symmetry on the breathing spectrum.
To this end, we initialized the system in the first excited state
having odd global parity as opposed to the ground state which is even.

For equal traps ($\eta=1$) we detected up to three modes.
First, at a weak majority-impurity interaction there is a beating.
One mode is of a majority type. 
It does a comparatively small contribution 
to the overall breathing dynamics while
its frequency is insensitive to $g_{AB}$ variations, albeit
depending on $g_A$ and $N_A$.
The mode becomes suppressed for a weakly attractive majority species
at positive $g_{AB}$.
The frequency of the second mode decreases monotonically
as a function of $g_{AB}$ until it bifurcates into two distinct frequencies.
One of them keeps decreasing.
Other parameters such as majority-component interaction strength $g_A$
or the number of majority atoms $N_A$ barely affect this frequency.
The corresponding mode is caused by an eigenstate of odd CM parity
and even global parity. 
The mode is equally represented in both components
while its existence depends on the entanglement.
The other emerging frequency bends evolving into a convex function
of the majority-impurity interaction.
The position of the minimum is very sensitive to $g_A$ and $N_A$.
The mode dominates the impurity motion.

By broadening the impurity trap ($\eta=0.51$)
only one mode can be excited
at weak $g_{AB}$.
It is of impurity type while
the corresponding frequency behaves as a convex function of $g_{AB}$.
With increasing impurity-majority interaction
a beating emerges.
At positive $g_{AB}$ the additional mode
is caused by the entanglement, while
at negative $g_{AB}$ both frequencies
are well-matched by the SMF ansatz.
For a tightly trapped impurity ($\eta=4$)
up to six frequencies can be observed,
though many more modes are in principal available.
The two impurity type modes are of high frequency but
are quickly fading away with increasing $g_{AB}$.
At strong $g_{AB}$ the impurity oscillations start imitating
the majority motion.
Among the lower frequency majority type modes 
there is one of major amplitude across all $g_{AB}$.
The corresponding frequency is coupling-insensitive
Among the modes emerging at finite $g_{AB}$ 
we emphasized the one caused by the entanglement.
With increasing $g_{AB}$ the mode amplitude is enhanced
while its frequency quickly decays.

Regarding the first excited state of odd global parity
as an initial state for the breathing dynamics, we found
the corresponding excitation spectrum to bear strong similarity
to the one of even parity ground state.
It can be understood as follows. 
Some of eigenstates lying in the odd parity subspace
have energies corresponding to the ones of even parity subspace
except for a constant energy shift, 
which is an integer number (in harmonic units)
corresponding to a center-of-mass excitation.
Nevertheless, there are also differences.
There is a slow-frequency mode equally represented in both components.
It is caused by the degeneracy of the first excited state.
The frequency starts at $\Omega=0$ and saturates towards $\Omega=1$
with increasing $g_{AB}$.
For a weakly repulsive majority species 
and at intermediate $g_{AB}$ 
we also observe several additional modes absent
in the ground state spectrum.

Overall, the few-body Bose-polaron breathing spectrum
has been studied 
and compared to the one of a particle-balanced
Bose-Bose mixture.
The quench protocol employed in this work
allowed to couple eigenstates of different CM parity (at $\eta=1$)
as opposed to a species-symmetric trap quench.
We excited a new kind of a breathing mode.
The eigenstate responsible for this mode has
odd CM parity ($\eta=1$) or can be traced back 
to an inter-component sloshing excitation at zero interactions ($\eta\neq1$).
The mode relies on the presence of entanglement,
while its frequency is a monotonically decreasing function of $g_{AB}$.
This opens the perspective
to study the relation between the mode amplitude
and the degree of entanglement, stored 
in the many-body composite state,
adding yet another item into the analysis toolbox
of breathing mode diagnostics.


\begin{acknowledgments}
	M.\ P.  gratefully  acknowledges  a  scholarship  
	of  the  Studienstiftung des deutschen Volkes.
\end{acknowledgments}

%

%

\bibliographystyle{apsrev4-2}

\begin{thebibliography}{72}%
	\makeatletter
	\providecommand \@ifxundefined [1]{%
		\@ifx{#1\undefined}
	}%
	\providecommand \@ifnum [1]{%
		\ifnum #1\expandafter \@firstoftwo
		\else \expandafter \@secondoftwo
		\fi
	}%
	\providecommand \@ifx [1]{%
		\ifx #1\expandafter \@firstoftwo
		\else \expandafter \@secondoftwo
		\fi
	}%
	\providecommand \natexlab [1]{#1}%
	\providecommand \enquote  [1]{``#1''}%
	\providecommand \bibnamefont  [1]{#1}%
	\providecommand \bibfnamefont [1]{#1}%
	\providecommand \citenamefont [1]{#1}%
	\providecommand \href@noop [0]{\@secondoftwo}%
	\providecommand \href [0]{\begingroup \@sanitize@url \@href}%
	\providecommand \@href[1]{\@@startlink{#1}\@@href}%
	\providecommand \@@href[1]{\endgroup#1\@@endlink}%
	\providecommand \@sanitize@url [0]{\catcode `\\12\catcode `\$12\catcode
		`\&12\catcode `\#12\catcode `\^12\catcode `\_12\catcode `\%12\relax}%
	\providecommand \@@startlink[1]{}%
	\providecommand \@@endlink[0]{}%
	\providecommand \url  [0]{\begingroup\@sanitize@url \@url }%
	\providecommand \@url [1]{\endgroup\@href {#1}{\urlprefix }}%
	\providecommand \urlprefix  [0]{URL }%
	\providecommand \Eprint [0]{\href }%
	\providecommand \doibase [0]{https://doi.org/}%
	\providecommand \selectlanguage [0]{\@gobble}%
	\providecommand \bibinfo  [0]{\@secondoftwo}%
	\providecommand \bibfield  [0]{\@secondoftwo}%
	\providecommand \translation [1]{[#1]}%
	\providecommand \BibitemOpen [0]{}%
	\providecommand \bibitemStop [0]{}%
	\providecommand \bibitemNoStop [0]{.\EOS\space}%
	\providecommand \EOS [0]{\spacefactor3000\relax}%
	\providecommand \BibitemShut  [1]{\csname bibitem#1\endcsname}%
	\let\auto@bib@innerbib\@empty
	\bibitem [{\citenamefont {Landau}(1933)}]{landau1933electron}%
	\BibitemOpen
	\bibfield  {author} {\bibinfo {author} {\bibfnamefont {L.~D.}\ \bibnamefont
			{Landau}},\ }\href@noop {} {\bibfield  {journal} {\bibinfo  {journal} {Phys.
				Z.}\ }\textbf {\bibinfo {volume} {3}},\ \bibinfo {pages} {664} (\bibinfo
		{year} {1933})}\BibitemShut {NoStop}%
	\bibitem [{\citenamefont {Pekar}(1946)}]{pekar1946local}%
	\BibitemOpen
	\bibfield  {author} {\bibinfo {author} {\bibfnamefont {S.}~\bibnamefont
			{Pekar}},\ }\href@noop {} {\bibfield  {journal} {\bibinfo  {journal} {Sov.
				Phys. JETP}\ }\textbf {\bibinfo {volume} {16}},\ \bibinfo {pages} {341}
		(\bibinfo {year} {1946})}\BibitemShut {NoStop}%
	\bibitem [{\citenamefont {Devreese}\ and\ \citenamefont
		{Alexandrov}(2009)}]{devreese2009frohlich}%
	\BibitemOpen
	\bibfield  {author} {\bibinfo {author} {\bibfnamefont {J.~T.}\ \bibnamefont
			{Devreese}}\ and\ \bibinfo {author} {\bibfnamefont {A.~S.}\ \bibnamefont
			{Alexandrov}},\ }\href@noop {} {\bibfield  {journal} {\bibinfo  {journal}
			{Rep. Prog. Phys.}\ }\textbf {\bibinfo {volume} {72}},\ \bibinfo {pages}
		{066501} (\bibinfo {year} {2009})}\BibitemShut {NoStop}%
	\bibitem [{\citenamefont {Alexandrov}\ and\ \citenamefont
		{Devreese}(2010)}]{alexandrov2010advances}%
	\BibitemOpen
	\bibfield  {author} {\bibinfo {author} {\bibfnamefont {A.~S.}\ \bibnamefont
			{Alexandrov}}\ and\ \bibinfo {author} {\bibfnamefont {J.~T.}\ \bibnamefont
			{Devreese}},\ }\href@noop {} {\emph {\bibinfo {title} {Advances in polaron
				physics}}},\ Vol.\ \bibinfo {volume} {159}\ (\bibinfo  {publisher}
	{Springer},\ \bibinfo {year} {2010})\BibitemShut {NoStop}%
	\bibitem [{\citenamefont {Mahan}(2013)}]{mahan2013many}%
	\BibitemOpen
	\bibfield  {author} {\bibinfo {author} {\bibfnamefont {G.~D.}\ \bibnamefont
			{Mahan}},\ }\href@noop {} {\emph {\bibinfo {title} {Many-particle physics}}}\
	(\bibinfo  {publisher} {Springer Science \& Business Media},\ \bibinfo {year}
	{2013})\BibitemShut {NoStop}%
	\bibitem [{\citenamefont {Anderson}\ \emph {et~al.}(1995)\citenamefont
		{Anderson}, \citenamefont {Ensher}, \citenamefont {Matthews}, \citenamefont
		{Wieman},\ and\ \citenamefont {Cornell}}]{anderson1995observation}%
	\BibitemOpen
	\bibfield  {author} {\bibinfo {author} {\bibfnamefont {M.~H.}\ \bibnamefont
			{Anderson}}, \bibinfo {author} {\bibfnamefont {J.~R.}\ \bibnamefont
			{Ensher}}, \bibinfo {author} {\bibfnamefont {M.~R.}\ \bibnamefont
			{Matthews}}, \bibinfo {author} {\bibfnamefont {C.~E.}\ \bibnamefont
			{Wieman}},\ and\ \bibinfo {author} {\bibfnamefont {E.~A.}\ \bibnamefont
			{Cornell}},\ }\href@noop {} {\bibfield  {journal} {\bibinfo  {journal}
			{Science}\ }\textbf {\bibinfo {volume} {269}},\ \bibinfo {pages} {198}
		(\bibinfo {year} {1995})}\BibitemShut {NoStop}%
	\bibitem [{\citenamefont {Davis}\ \emph {et~al.}(1995)\citenamefont {Davis},
		\citenamefont {Mewes}, \citenamefont {Andrews}, \citenamefont {van Druten},
		\citenamefont {Durfee}, \citenamefont {Kurn},\ and\ \citenamefont
		{Ketterle}}]{davis1995bose}%
	\BibitemOpen
	\bibfield  {author} {\bibinfo {author} {\bibfnamefont {K.~B.}\ \bibnamefont
			{Davis}}, \bibinfo {author} {\bibfnamefont {M.-O.}\ \bibnamefont {Mewes}},
		\bibinfo {author} {\bibfnamefont {M.~R.}\ \bibnamefont {Andrews}}, \bibinfo
		{author} {\bibfnamefont {N.~J.}\ \bibnamefont {van Druten}}, \bibinfo
		{author} {\bibfnamefont {D.~S.}\ \bibnamefont {Durfee}}, \bibinfo {author}
		{\bibfnamefont {D.}~\bibnamefont {Kurn}},\ and\ \bibinfo {author}
		{\bibfnamefont {W.}~\bibnamefont {Ketterle}},\ }\href@noop {} {\bibfield
		{journal} {\bibinfo  {journal} {Phys. Rev. Lett.}\ }\textbf {\bibinfo
			{volume} {75}},\ \bibinfo {pages} {3969} (\bibinfo {year}
		{1995})}\BibitemShut {NoStop}%
	\bibitem [{\citenamefont {Bloch}\ \emph {et~al.}(2008)\citenamefont {Bloch},
		\citenamefont {Dalibard},\ and\ \citenamefont {Zwerger}}]{bloch2008many}%
	\BibitemOpen
	\bibfield  {author} {\bibinfo {author} {\bibfnamefont {I.}~\bibnamefont
			{Bloch}}, \bibinfo {author} {\bibfnamefont {J.}~\bibnamefont {Dalibard}},\
		and\ \bibinfo {author} {\bibfnamefont {W.}~\bibnamefont {Zwerger}},\
	}\href@noop {} {\bibfield  {journal} {\bibinfo  {journal} {Rev. Mod. Phys.}\
		}\textbf {\bibinfo {volume} {80}},\ \bibinfo {pages} {885} (\bibinfo {year}
		{2008})}\BibitemShut {NoStop}%
	\bibitem [{\citenamefont {Myatt}\ \emph {et~al.}(1997)\citenamefont {Myatt},
		\citenamefont {Burt}, \citenamefont {Ghrist}, \citenamefont {Cornell},\ and\
		\citenamefont {Wieman}}]{myatt1997production}%
	\BibitemOpen
	\bibfield  {author} {\bibinfo {author} {\bibfnamefont {C.}~\bibnamefont
			{Myatt}}, \bibinfo {author} {\bibfnamefont {E.}~\bibnamefont {Burt}},
		\bibinfo {author} {\bibfnamefont {R.}~\bibnamefont {Ghrist}}, \bibinfo
		{author} {\bibfnamefont {E.~A.}\ \bibnamefont {Cornell}},\ and\ \bibinfo
		{author} {\bibfnamefont {C.}~\bibnamefont {Wieman}},\ }\href@noop {}
	{\bibfield  {journal} {\bibinfo  {journal} {Phys. Rev. Lett.}\ }\textbf
		{\bibinfo {volume} {78}},\ \bibinfo {pages} {586} (\bibinfo {year}
		{1997})}\BibitemShut {NoStop}%
	\bibitem [{\citenamefont {Blume}(2012)}]{blume2012few}%
	\BibitemOpen
	\bibfield  {author} {\bibinfo {author} {\bibfnamefont {D.}~\bibnamefont
			{Blume}},\ }\href@noop {} {\bibfield  {journal} {\bibinfo  {journal} {Rep.
				Prog. Phys.}\ }\textbf {\bibinfo {volume} {75}},\ \bibinfo {pages} {046401}
		(\bibinfo {year} {2012})}\BibitemShut {NoStop}%
	\bibitem [{\citenamefont {Palzer}\ \emph {et~al.}(2009)\citenamefont {Palzer},
		\citenamefont {Zipkes}, \citenamefont {Sias},\ and\ \citenamefont
		{K{\"o}hl}}]{palzer2009quantum}%
	\BibitemOpen
	\bibfield  {author} {\bibinfo {author} {\bibfnamefont {S.}~\bibnamefont
			{Palzer}}, \bibinfo {author} {\bibfnamefont {C.}~\bibnamefont {Zipkes}},
		\bibinfo {author} {\bibfnamefont {C.}~\bibnamefont {Sias}},\ and\ \bibinfo
		{author} {\bibfnamefont {M.}~\bibnamefont {K{\"o}hl}},\ }\href@noop {}
	{\bibfield  {journal} {\bibinfo  {journal} {Phys. Rev. Lett.}\ }\textbf
		{\bibinfo {volume} {103}},\ \bibinfo {pages} {150601} (\bibinfo {year}
		{2009})}\BibitemShut {NoStop}%
	\bibitem [{\citenamefont {Catani}\ \emph {et~al.}(2012)\citenamefont {Catani},
		\citenamefont {Lamporesi}, \citenamefont {Naik}, \citenamefont {Gring},
		\citenamefont {Inguscio}, \citenamefont {Minardi}, \citenamefont {Kantian},\
		and\ \citenamefont {Giamarchi}}]{catani2012quantum}%
	\BibitemOpen
	\bibfield  {author} {\bibinfo {author} {\bibfnamefont {J.}~\bibnamefont
			{Catani}}, \bibinfo {author} {\bibfnamefont {G.}~\bibnamefont {Lamporesi}},
		\bibinfo {author} {\bibfnamefont {D.}~\bibnamefont {Naik}}, \bibinfo {author}
		{\bibfnamefont {M.}~\bibnamefont {Gring}}, \bibinfo {author} {\bibfnamefont
			{M.}~\bibnamefont {Inguscio}}, \bibinfo {author} {\bibfnamefont
			{F.}~\bibnamefont {Minardi}}, \bibinfo {author} {\bibfnamefont
			{A.}~\bibnamefont {Kantian}},\ and\ \bibinfo {author} {\bibfnamefont
			{T.}~\bibnamefont {Giamarchi}},\ }\href@noop {} {\bibfield  {journal}
		{\bibinfo  {journal} {Phys. Rev. A}\ }\textbf {\bibinfo {volume} {85}},\
		\bibinfo {pages} {023623} (\bibinfo {year} {2012})}\BibitemShut {NoStop}%
	\bibitem [{\citenamefont {Fukuhara}\ \emph {et~al.}(2013)\citenamefont
		{Fukuhara}, \citenamefont {Kantian}, \citenamefont {Endres}, \citenamefont
		{Cheneau}, \citenamefont {Schau{\ss}}, \citenamefont {Hild}, \citenamefont
		{Bellem}, \citenamefont {Schollw{\"o}ck}, \citenamefont {Giamarchi},
		\citenamefont {Gross} \emph {et~al.}}]{fukuhara2013quantum}%
	\BibitemOpen
	\bibfield  {author} {\bibinfo {author} {\bibfnamefont {T.}~\bibnamefont
			{Fukuhara}}, \bibinfo {author} {\bibfnamefont {A.}~\bibnamefont {Kantian}},
		\bibinfo {author} {\bibfnamefont {M.}~\bibnamefont {Endres}}, \bibinfo
		{author} {\bibfnamefont {M.}~\bibnamefont {Cheneau}}, \bibinfo {author}
		{\bibfnamefont {P.}~\bibnamefont {Schau{\ss}}}, \bibinfo {author}
		{\bibfnamefont {S.}~\bibnamefont {Hild}}, \bibinfo {author} {\bibfnamefont
			{D.}~\bibnamefont {Bellem}}, \bibinfo {author} {\bibfnamefont
			{U.}~\bibnamefont {Schollw{\"o}ck}}, \bibinfo {author} {\bibfnamefont
			{T.}~\bibnamefont {Giamarchi}}, \bibinfo {author} {\bibfnamefont
			{C.}~\bibnamefont {Gross}}, \emph {et~al.},\ }\href@noop {} {\bibfield
		{journal} {\bibinfo  {journal} {Nat. Phys.}\ }\textbf {\bibinfo {volume}
			{9}},\ \bibinfo {pages} {235} (\bibinfo {year} {2013})}\BibitemShut {NoStop}%
	\bibitem [{\citenamefont {Spethmann}\ \emph {et~al.}(2012)\citenamefont
		{Spethmann}, \citenamefont {Kindermann}, \citenamefont {John}, \citenamefont
		{Weber}, \citenamefont {Meschede},\ and\ \citenamefont
		{Widera}}]{spethmann2012dynamics}%
	\BibitemOpen
	\bibfield  {author} {\bibinfo {author} {\bibfnamefont {N.}~\bibnamefont
			{Spethmann}}, \bibinfo {author} {\bibfnamefont {F.}~\bibnamefont
			{Kindermann}}, \bibinfo {author} {\bibfnamefont {S.}~\bibnamefont {John}},
		\bibinfo {author} {\bibfnamefont {C.}~\bibnamefont {Weber}}, \bibinfo
		{author} {\bibfnamefont {D.}~\bibnamefont {Meschede}},\ and\ \bibinfo
		{author} {\bibfnamefont {A.}~\bibnamefont {Widera}},\ }\href@noop {}
	{\bibfield  {journal} {\bibinfo  {journal} {Phys. Rev. Lett.}\ }\textbf
		{\bibinfo {volume} {109}},\ \bibinfo {pages} {235301} (\bibinfo {year}
		{2012})}\BibitemShut {NoStop}%
	\bibitem [{\citenamefont {J{\o}rgensen}\ \emph {et~al.}(2016)\citenamefont
		{J{\o}rgensen}, \citenamefont {Wacker}, \citenamefont {Skalmstang},
		\citenamefont {Parish}, \citenamefont {Levinsen}, \citenamefont
		{Christensen}, \citenamefont {Bruun},\ and\ \citenamefont
		{Arlt}}]{jorgensen2016observation}%
	\BibitemOpen
	\bibfield  {author} {\bibinfo {author} {\bibfnamefont {N.~B.}\ \bibnamefont
			{J{\o}rgensen}}, \bibinfo {author} {\bibfnamefont {L.}~\bibnamefont
			{Wacker}}, \bibinfo {author} {\bibfnamefont {K.~T.}\ \bibnamefont
			{Skalmstang}}, \bibinfo {author} {\bibfnamefont {M.~M.}\ \bibnamefont
			{Parish}}, \bibinfo {author} {\bibfnamefont {J.}~\bibnamefont {Levinsen}},
		\bibinfo {author} {\bibfnamefont {R.~S.}\ \bibnamefont {Christensen}},
		\bibinfo {author} {\bibfnamefont {G.~M.}\ \bibnamefont {Bruun}},\ and\
		\bibinfo {author} {\bibfnamefont {J.~J.}\ \bibnamefont {Arlt}},\ }\href@noop
	{} {\bibfield  {journal} {\bibinfo  {journal} {Phys. Rev. Lett.}\ }\textbf
		{\bibinfo {volume} {117}},\ \bibinfo {pages} {055302} (\bibinfo {year}
		{2016})}\BibitemShut {NoStop}%
	\bibitem [{\citenamefont {Meinert}\ \emph {et~al.}(2017)\citenamefont
		{Meinert}, \citenamefont {Knap}, \citenamefont {Kirilov}, \citenamefont
		{Jag-Lauber}, \citenamefont {Zvonarev}, \citenamefont {Demler},\ and\
		\citenamefont {N{\"a}gerl}}]{meinert2017bloch}%
	\BibitemOpen
	\bibfield  {author} {\bibinfo {author} {\bibfnamefont {F.}~\bibnamefont
			{Meinert}}, \bibinfo {author} {\bibfnamefont {M.}~\bibnamefont {Knap}},
		\bibinfo {author} {\bibfnamefont {E.}~\bibnamefont {Kirilov}}, \bibinfo
		{author} {\bibfnamefont {K.}~\bibnamefont {Jag-Lauber}}, \bibinfo {author}
		{\bibfnamefont {M.~B.}\ \bibnamefont {Zvonarev}}, \bibinfo {author}
		{\bibfnamefont {E.}~\bibnamefont {Demler}},\ and\ \bibinfo {author}
		{\bibfnamefont {H.-C.}\ \bibnamefont {N{\"a}gerl}},\ }\href@noop {}
	{\bibfield  {journal} {\bibinfo  {journal} {Science}\ }\textbf {\bibinfo
			{volume} {356}},\ \bibinfo {pages} {945} (\bibinfo {year}
		{2017})}\BibitemShut {NoStop}%
	\bibitem [{\citenamefont {Nascimbene}\ \emph {et~al.}(2009)\citenamefont
		{Nascimbene}, \citenamefont {Navon}, \citenamefont {Jiang}, \citenamefont
		{Tarruell}, \citenamefont {Teichmann}, \citenamefont {Mckeever},
		\citenamefont {Chevy},\ and\ \citenamefont
		{Salomon}}]{nascimbene2009collective}%
	\BibitemOpen
	\bibfield  {author} {\bibinfo {author} {\bibfnamefont {S.}~\bibnamefont
			{Nascimbene}}, \bibinfo {author} {\bibfnamefont {N.}~\bibnamefont {Navon}},
		\bibinfo {author} {\bibfnamefont {K.}~\bibnamefont {Jiang}}, \bibinfo
		{author} {\bibfnamefont {L.}~\bibnamefont {Tarruell}}, \bibinfo {author}
		{\bibfnamefont {M.}~\bibnamefont {Teichmann}}, \bibinfo {author}
		{\bibfnamefont {J.}~\bibnamefont {Mckeever}}, \bibinfo {author}
		{\bibfnamefont {F.}~\bibnamefont {Chevy}},\ and\ \bibinfo {author}
		{\bibfnamefont {C.}~\bibnamefont {Salomon}},\ }\href@noop {} {\bibfield
		{journal} {\bibinfo  {journal} {Phys. Rev. Lett.}\ }\textbf {\bibinfo
			{volume} {103}},\ \bibinfo {pages} {170402} (\bibinfo {year}
		{2009})}\BibitemShut {NoStop}%
	\bibitem [{\citenamefont {Schirotzek}\ \emph {et~al.}(2009)\citenamefont
		{Schirotzek}, \citenamefont {Wu}, \citenamefont {Sommer},\ and\ \citenamefont
		{Zwierlein}}]{schirotzek2009observation}%
	\BibitemOpen
	\bibfield  {author} {\bibinfo {author} {\bibfnamefont {A.}~\bibnamefont
			{Schirotzek}}, \bibinfo {author} {\bibfnamefont {C.-H.}\ \bibnamefont {Wu}},
		\bibinfo {author} {\bibfnamefont {A.}~\bibnamefont {Sommer}},\ and\ \bibinfo
		{author} {\bibfnamefont {M.~W.}\ \bibnamefont {Zwierlein}},\ }\href@noop {}
	{\bibfield  {journal} {\bibinfo  {journal} {Phys. Rev. Lett.}\ }\textbf
		{\bibinfo {volume} {102}},\ \bibinfo {pages} {230402} (\bibinfo {year}
		{2009})}\BibitemShut {NoStop}%
	\bibitem [{\citenamefont {Koschorreck}\ \emph {et~al.}(2012)\citenamefont
		{Koschorreck}, \citenamefont {Pertot}, \citenamefont {Vogt}, \citenamefont
		{Fr{\"o}hlich}, \citenamefont {Feld},\ and\ \citenamefont
		{K{\"o}hl}}]{koschorreck2012attractive}%
	\BibitemOpen
	\bibfield  {author} {\bibinfo {author} {\bibfnamefont {M.}~\bibnamefont
			{Koschorreck}}, \bibinfo {author} {\bibfnamefont {D.}~\bibnamefont {Pertot}},
		\bibinfo {author} {\bibfnamefont {E.}~\bibnamefont {Vogt}}, \bibinfo {author}
		{\bibfnamefont {B.}~\bibnamefont {Fr{\"o}hlich}}, \bibinfo {author}
		{\bibfnamefont {M.}~\bibnamefont {Feld}},\ and\ \bibinfo {author}
		{\bibfnamefont {M.}~\bibnamefont {K{\"o}hl}},\ }\href@noop {} {\bibfield
		{journal} {\bibinfo  {journal} {Nature}\ }\textbf {\bibinfo {volume} {485}},\
		\bibinfo {pages} {619} (\bibinfo {year} {2012})}\BibitemShut {NoStop}%
	\bibitem [{\citenamefont {Kohstall}\ \emph {et~al.}(2012)\citenamefont
		{Kohstall}, \citenamefont {Zaccanti}, \citenamefont {Jag}, \citenamefont
		{Trenkwalder}, \citenamefont {Massignan}, \citenamefont {Bruun},
		\citenamefont {Schreck},\ and\ \citenamefont
		{Grimm}}]{kohstall2012metastability}%
	\BibitemOpen
	\bibfield  {author} {\bibinfo {author} {\bibfnamefont {C.}~\bibnamefont
			{Kohstall}}, \bibinfo {author} {\bibfnamefont {M.}~\bibnamefont {Zaccanti}},
		\bibinfo {author} {\bibfnamefont {M.}~\bibnamefont {Jag}}, \bibinfo {author}
		{\bibfnamefont {A.}~\bibnamefont {Trenkwalder}}, \bibinfo {author}
		{\bibfnamefont {P.}~\bibnamefont {Massignan}}, \bibinfo {author}
		{\bibfnamefont {G.~M.}\ \bibnamefont {Bruun}}, \bibinfo {author}
		{\bibfnamefont {F.}~\bibnamefont {Schreck}},\ and\ \bibinfo {author}
		{\bibfnamefont {R.}~\bibnamefont {Grimm}},\ }\href@noop {} {\bibfield
		{journal} {\bibinfo  {journal} {Nature}\ }\textbf {\bibinfo {volume} {485}},\
		\bibinfo {pages} {615} (\bibinfo {year} {2012})}\BibitemShut {NoStop}%
	\bibitem [{\citenamefont {Cetina}\ \emph {et~al.}(2015)\citenamefont {Cetina},
		\citenamefont {Jag}, \citenamefont {Lous}, \citenamefont {Walraven},
		\citenamefont {Grimm}, \citenamefont {Christensen},\ and\ \citenamefont
		{Bruun}}]{cetina2015decoherence}%
	\BibitemOpen
	\bibfield  {author} {\bibinfo {author} {\bibfnamefont {M.}~\bibnamefont
			{Cetina}}, \bibinfo {author} {\bibfnamefont {M.}~\bibnamefont {Jag}},
		\bibinfo {author} {\bibfnamefont {R.~S.}\ \bibnamefont {Lous}}, \bibinfo
		{author} {\bibfnamefont {J.~T.}\ \bibnamefont {Walraven}}, \bibinfo {author}
		{\bibfnamefont {R.}~\bibnamefont {Grimm}}, \bibinfo {author} {\bibfnamefont
			{R.~S.}\ \bibnamefont {Christensen}},\ and\ \bibinfo {author} {\bibfnamefont
			{G.~M.}\ \bibnamefont {Bruun}},\ }\href@noop {} {\bibfield  {journal}
		{\bibinfo  {journal} {Phys. Rev. Lett.}\ }\textbf {\bibinfo {volume} {115}},\
		\bibinfo {pages} {135302} (\bibinfo {year} {2015})}\BibitemShut {NoStop}%
	\bibitem [{\citenamefont {Grusdt}\ and\ \citenamefont
		{Demler}(2015)}]{grusdt2015new}%
	\BibitemOpen
	\bibfield  {author} {\bibinfo {author} {\bibfnamefont {F.}~\bibnamefont
			{Grusdt}}\ and\ \bibinfo {author} {\bibfnamefont {E.}~\bibnamefont
			{Demler}},\ }\href@noop {} {\bibfield  {journal} {\bibinfo  {journal}
			{Quantum Matter at Ultralow Temperatures}\ }\textbf {\bibinfo {volume}
			{191}},\ \bibinfo {pages} {325} (\bibinfo {year} {2015})}\BibitemShut
	{NoStop}%
	\bibitem [{\citenamefont {Chevy}\ and\ \citenamefont
		{Mora}(2010)}]{chevy2010ultra}%
	\BibitemOpen
	\bibfield  {author} {\bibinfo {author} {\bibfnamefont {F.}~\bibnamefont
			{Chevy}}\ and\ \bibinfo {author} {\bibfnamefont {C.}~\bibnamefont {Mora}},\
	}\href@noop {} {\bibfield  {journal} {\bibinfo  {journal} {Rep. Prog. Phys.}\
		}\textbf {\bibinfo {volume} {73}},\ \bibinfo {pages} {112401} (\bibinfo
		{year} {2010})}\BibitemShut {NoStop}%
	\bibitem [{\citenamefont {Massignan}\ \emph {et~al.}(2014)\citenamefont
		{Massignan}, \citenamefont {Zaccanti},\ and\ \citenamefont
		{Bruun}}]{massignan2014polarons}%
	\BibitemOpen
	\bibfield  {author} {\bibinfo {author} {\bibfnamefont {P.}~\bibnamefont
			{Massignan}}, \bibinfo {author} {\bibfnamefont {M.}~\bibnamefont
			{Zaccanti}},\ and\ \bibinfo {author} {\bibfnamefont {G.~M.}\ \bibnamefont
			{Bruun}},\ }\href@noop {} {\bibfield  {journal} {\bibinfo  {journal} {Rep.
				Prog. Phys.}\ }\textbf {\bibinfo {volume} {77}},\ \bibinfo {pages} {034401}
		(\bibinfo {year} {2014})}\BibitemShut {NoStop}%
	\bibitem [{\citenamefont {Fr{\"o}hlich}(1954)}]{frohlich1954electrons}%
	\BibitemOpen
	\bibfield  {author} {\bibinfo {author} {\bibfnamefont {H.}~\bibnamefont
			{Fr{\"o}hlich}},\ }\href@noop {} {\bibfield  {journal} {\bibinfo  {journal}
			{Adv. Phys.}\ }\textbf {\bibinfo {volume} {3}},\ \bibinfo {pages} {325}
		(\bibinfo {year} {1954})}\BibitemShut {NoStop}%
	\bibitem [{\citenamefont {Chin}\ \emph {et~al.}(2010)\citenamefont {Chin},
		\citenamefont {Grimm}, \citenamefont {Julienne},\ and\ \citenamefont
		{Tiesinga}}]{chin2010feshbach}%
	\BibitemOpen
	\bibfield  {author} {\bibinfo {author} {\bibfnamefont {C.}~\bibnamefont
			{Chin}}, \bibinfo {author} {\bibfnamefont {R.}~\bibnamefont {Grimm}},
		\bibinfo {author} {\bibfnamefont {P.}~\bibnamefont {Julienne}},\ and\
		\bibinfo {author} {\bibfnamefont {E.}~\bibnamefont {Tiesinga}},\ }\href@noop
	{} {\bibfield  {journal} {\bibinfo  {journal} {Rev. Mod. Phys.}\ }\textbf
		{\bibinfo {volume} {82}},\ \bibinfo {pages} {1225} (\bibinfo {year}
		{2010})}\BibitemShut {NoStop}%
	\bibitem [{\citenamefont {K{\"o}hler}\ \emph {et~al.}(2006)\citenamefont
		{K{\"o}hler}, \citenamefont {G{\'o}ral},\ and\ \citenamefont
		{Julienne}}]{kohler2006production}%
	\BibitemOpen
	\bibfield  {author} {\bibinfo {author} {\bibfnamefont {T.}~\bibnamefont
			{K{\"o}hler}}, \bibinfo {author} {\bibfnamefont {K.}~\bibnamefont
			{G{\'o}ral}},\ and\ \bibinfo {author} {\bibfnamefont {P.~S.}\ \bibnamefont
			{Julienne}},\ }\href@noop {} {\bibfield  {journal} {\bibinfo  {journal} {Rev.
				Mod. Phys.}\ }\textbf {\bibinfo {volume} {78}},\ \bibinfo {pages} {1311}
		(\bibinfo {year} {2006})}\BibitemShut {NoStop}%
	\bibitem [{\citenamefont {Volosniev}\ and\ \citenamefont
		{Hammer}(2017)}]{volosniev2017analytical}%
	\BibitemOpen
	\bibfield  {author} {\bibinfo {author} {\bibfnamefont {A.}~\bibnamefont
			{Volosniev}}\ and\ \bibinfo {author} {\bibfnamefont {H.-W.}\ \bibnamefont
			{Hammer}},\ }\href@noop {} {\bibfield  {journal} {\bibinfo  {journal} {Phys.
				Rev. A}\ }\textbf {\bibinfo {volume} {96}},\ \bibinfo {pages} {031601}
		(\bibinfo {year} {2017})}\BibitemShut {NoStop}%
	\bibitem [{\citenamefont {Grusdt}\ \emph {et~al.}(2017)\citenamefont {Grusdt},
		\citenamefont {Astrakharchik},\ and\ \citenamefont
		{Demler}}]{grusdt2017bose}%
	\BibitemOpen
	\bibfield  {author} {\bibinfo {author} {\bibfnamefont {F.}~\bibnamefont
			{Grusdt}}, \bibinfo {author} {\bibfnamefont {G.~E.}\ \bibnamefont
			{Astrakharchik}},\ and\ \bibinfo {author} {\bibfnamefont {E.}~\bibnamefont
			{Demler}},\ }\href@noop {} {\bibfield  {journal} {\bibinfo  {journal} {New J.
				Phys.}\ }\textbf {\bibinfo {volume} {19}},\ \bibinfo {pages} {103035}
		(\bibinfo {year} {2017})}\BibitemShut {NoStop}%
	\bibitem [{\citenamefont {Grusdt}\ \emph {et~al.}(2018)\citenamefont {Grusdt},
		\citenamefont {Seetharam}, \citenamefont {Shchadilova},\ and\ \citenamefont
		{Demler}}]{grusdt2018strong}%
	\BibitemOpen
	\bibfield  {author} {\bibinfo {author} {\bibfnamefont {F.}~\bibnamefont
			{Grusdt}}, \bibinfo {author} {\bibfnamefont {K.}~\bibnamefont {Seetharam}},
		\bibinfo {author} {\bibfnamefont {Y.}~\bibnamefont {Shchadilova}},\ and\
		\bibinfo {author} {\bibfnamefont {E.}~\bibnamefont {Demler}},\ }\href@noop {}
	{\bibfield  {journal} {\bibinfo  {journal} {Phys. Rev. A}\ }\textbf {\bibinfo
			{volume} {97}},\ \bibinfo {pages} {033612} (\bibinfo {year}
		{2018})}\BibitemShut {NoStop}%
	\bibitem [{\citenamefont {Kain}\ and\ \citenamefont
		{Ling}(2018)}]{kain2018analytical}%
	\BibitemOpen
	\bibfield  {author} {\bibinfo {author} {\bibfnamefont {B.}~\bibnamefont
			{Kain}}\ and\ \bibinfo {author} {\bibfnamefont {H.~Y.}\ \bibnamefont
			{Ling}},\ }\href@noop {} {\bibfield  {journal} {\bibinfo  {journal} {Phys.
				Rev. A}\ }\textbf {\bibinfo {volume} {98}},\ \bibinfo {pages} {033610}
		(\bibinfo {year} {2018})}\BibitemShut {NoStop}%
	\bibitem [{\citenamefont {Drescher}\ \emph {et~al.}(2019)\citenamefont
		{Drescher}, \citenamefont {Salmhofer},\ and\ \citenamefont
		{Enss}}]{drescher2019real}%
	\BibitemOpen
	\bibfield  {author} {\bibinfo {author} {\bibfnamefont {M.}~\bibnamefont
			{Drescher}}, \bibinfo {author} {\bibfnamefont {M.}~\bibnamefont
			{Salmhofer}},\ and\ \bibinfo {author} {\bibfnamefont {T.}~\bibnamefont
			{Enss}},\ }\href@noop {} {\bibfield  {journal} {\bibinfo  {journal} {Phys.
				Rev. A}\ }\textbf {\bibinfo {volume} {99}},\ \bibinfo {pages} {023601}
		(\bibinfo {year} {2019})}\BibitemShut {NoStop}%
	\bibitem [{\citenamefont {Jager}\ \emph {et~al.}(2020)\citenamefont {Jager},
		\citenamefont {Barnett}, \citenamefont {Will},\ and\ \citenamefont
		{Fleischhauer}}]{jager2020strong}%
	\BibitemOpen
	\bibfield  {author} {\bibinfo {author} {\bibfnamefont {J.}~\bibnamefont
			{Jager}}, \bibinfo {author} {\bibfnamefont {R.}~\bibnamefont {Barnett}},
		\bibinfo {author} {\bibfnamefont {M.}~\bibnamefont {Will}},\ and\ \bibinfo
		{author} {\bibfnamefont {M.}~\bibnamefont {Fleischhauer}},\ }\href@noop {}
	{\bibfield  {journal} {\bibinfo  {journal} {Phys. Rev. Res.}\ }\textbf
		{\bibinfo {volume} {2}},\ \bibinfo {pages} {033142} (\bibinfo {year}
		{2020})}\BibitemShut {NoStop}%
	\bibitem [{\citenamefont {Drescher}\ \emph {et~al.}(2020)\citenamefont
		{Drescher}, \citenamefont {Salmhofer},\ and\ \citenamefont
		{Enss}}]{drescher2020theory}%
	\BibitemOpen
	\bibfield  {author} {\bibinfo {author} {\bibfnamefont {M.}~\bibnamefont
			{Drescher}}, \bibinfo {author} {\bibfnamefont {M.}~\bibnamefont
			{Salmhofer}},\ and\ \bibinfo {author} {\bibfnamefont {T.}~\bibnamefont
			{Enss}},\ }\href@noop {} {\bibfield  {journal} {\bibinfo  {journal} {Phys.
				Rev. Res.}\ }\textbf {\bibinfo {volume} {2}},\ \bibinfo {pages} {032011}
		(\bibinfo {year} {2020})}\BibitemShut {NoStop}%
	\bibitem [{\citenamefont {Giamarchi}(2003)}]{giamarchi2003quantum}%
	\BibitemOpen
	\bibfield  {author} {\bibinfo {author} {\bibfnamefont {T.}~\bibnamefont
			{Giamarchi}},\ }\href@noop {} {\emph {\bibinfo {title} {Quantum physics in
				one dimension}}},\ Vol.\ \bibinfo {volume} {121}\ (\bibinfo  {publisher}
	{Clarendon press},\ \bibinfo {year} {2003})\BibitemShut {NoStop}%
	\bibitem [{\citenamefont {Olshanii}(1998)}]{olshanii1998atomic}%
	\BibitemOpen
	\bibfield  {author} {\bibinfo {author} {\bibfnamefont {M.}~\bibnamefont
			{Olshanii}},\ }\href@noop {} {\bibfield  {journal} {\bibinfo  {journal}
			{Phys. Rev. Lett.}\ }\textbf {\bibinfo {volume} {81}},\ \bibinfo {pages}
		{938} (\bibinfo {year} {1998})}\BibitemShut {NoStop}%
	\bibitem [{\citenamefont {Bergeman}\ \emph {et~al.}(2003)\citenamefont
		{Bergeman}, \citenamefont {Moore},\ and\ \citenamefont
		{Olshanii}}]{bergeman2003atom}%
	\BibitemOpen
	\bibfield  {author} {\bibinfo {author} {\bibfnamefont {T.}~\bibnamefont
			{Bergeman}}, \bibinfo {author} {\bibfnamefont {M.}~\bibnamefont {Moore}},\
		and\ \bibinfo {author} {\bibfnamefont {M.}~\bibnamefont {Olshanii}},\
	}\href@noop {} {\bibfield  {journal} {\bibinfo  {journal} {Phys. Rev. Lett.}\
		}\textbf {\bibinfo {volume} {91}},\ \bibinfo {pages} {163201} (\bibinfo
		{year} {2003})}\BibitemShut {NoStop}%
	\bibitem [{\citenamefont {Haller}\ \emph {et~al.}(2010)\citenamefont {Haller},
		\citenamefont {Mark}, \citenamefont {Hart}, \citenamefont {Danzl},
		\citenamefont {Reichs{\"o}llner}, \citenamefont {Melezhik}, \citenamefont
		{Schmelcher},\ and\ \citenamefont {N{\"a}gerl}}]{haller2010confinement}%
	\BibitemOpen
	\bibfield  {author} {\bibinfo {author} {\bibfnamefont {E.}~\bibnamefont
			{Haller}}, \bibinfo {author} {\bibfnamefont {M.~J.}\ \bibnamefont {Mark}},
		\bibinfo {author} {\bibfnamefont {R.}~\bibnamefont {Hart}}, \bibinfo {author}
		{\bibfnamefont {J.~G.}\ \bibnamefont {Danzl}}, \bibinfo {author}
		{\bibfnamefont {L.}~\bibnamefont {Reichs{\"o}llner}}, \bibinfo {author}
		{\bibfnamefont {V.}~\bibnamefont {Melezhik}}, \bibinfo {author}
		{\bibfnamefont {P.}~\bibnamefont {Schmelcher}},\ and\ \bibinfo {author}
		{\bibfnamefont {H.-C.}\ \bibnamefont {N{\"a}gerl}},\ }\href@noop {}
	{\bibfield  {journal} {\bibinfo  {journal} {Phys. Rev. Lett.}\ }\textbf
		{\bibinfo {volume} {104}},\ \bibinfo {pages} {153203} (\bibinfo {year}
		{2010})}\BibitemShut {NoStop}%
	\bibitem [{\citenamefont {Girardeau}(1960)}]{girardeau1960relationship}%
	\BibitemOpen
	\bibfield  {author} {\bibinfo {author} {\bibfnamefont {M.}~\bibnamefont
			{Girardeau}},\ }\href@noop {} {\bibfield  {journal} {\bibinfo  {journal} {J.
				Math. Phys.}\ }\textbf {\bibinfo {volume} {1}},\ \bibinfo {pages} {516}
		(\bibinfo {year} {1960})}\BibitemShut {NoStop}%
	\bibitem [{\citenamefont {Kinoshita}\ \emph {et~al.}(2004)\citenamefont
		{Kinoshita}, \citenamefont {Wenger},\ and\ \citenamefont
		{Weiss}}]{kinoshita2004observation}%
	\BibitemOpen
	\bibfield  {author} {\bibinfo {author} {\bibfnamefont {T.}~\bibnamefont
			{Kinoshita}}, \bibinfo {author} {\bibfnamefont {T.}~\bibnamefont {Wenger}},\
		and\ \bibinfo {author} {\bibfnamefont {D.~S.}\ \bibnamefont {Weiss}},\
	}\href@noop {} {\bibfield  {journal} {\bibinfo  {journal} {Science}\ }\textbf
		{\bibinfo {volume} {305}},\ \bibinfo {pages} {1125} (\bibinfo {year}
		{2004})}\BibitemShut {NoStop}%
	\bibitem [{\citenamefont {Paredes}\ \emph {et~al.}(2004)\citenamefont
		{Paredes}, \citenamefont {Widera}, \citenamefont {Murg}, \citenamefont
		{Mandel}, \citenamefont {F{\"o}lling}, \citenamefont {Cirac}, \citenamefont
		{Shlyapnikov}, \citenamefont {H{\"a}nsch},\ and\ \citenamefont
		{Bloch}}]{paredes2004tonks}%
	\BibitemOpen
	\bibfield  {author} {\bibinfo {author} {\bibfnamefont {B.}~\bibnamefont
			{Paredes}}, \bibinfo {author} {\bibfnamefont {A.}~\bibnamefont {Widera}},
		\bibinfo {author} {\bibfnamefont {V.}~\bibnamefont {Murg}}, \bibinfo {author}
		{\bibfnamefont {O.}~\bibnamefont {Mandel}}, \bibinfo {author} {\bibfnamefont
			{S.}~\bibnamefont {F{\"o}lling}}, \bibinfo {author} {\bibfnamefont
			{I.}~\bibnamefont {Cirac}}, \bibinfo {author} {\bibfnamefont {G.~V.}\
			\bibnamefont {Shlyapnikov}}, \bibinfo {author} {\bibfnamefont {T.~W.}\
			\bibnamefont {H{\"a}nsch}},\ and\ \bibinfo {author} {\bibfnamefont
			{I.}~\bibnamefont {Bloch}},\ }\href@noop {} {\bibfield  {journal} {\bibinfo
			{journal} {Nature}\ }\textbf {\bibinfo {volume} {429}},\ \bibinfo {pages}
		{277} (\bibinfo {year} {2004})}\BibitemShut {NoStop}%
	\bibitem [{\citenamefont {Sowi{\'n}ski}\ and\ \citenamefont
		{Garc{\'\i}a-March}(2019)}]{sowinski2019one}%
	\BibitemOpen
	\bibfield  {author} {\bibinfo {author} {\bibfnamefont {T.}~\bibnamefont
			{Sowi{\'n}ski}}\ and\ \bibinfo {author} {\bibfnamefont {M.~{\'A}.}\
			\bibnamefont {Garc{\'\i}a-March}},\ }\href@noop {} {\bibfield  {journal}
		{\bibinfo  {journal} {Rep. Prog. Phys.}\ }\textbf {\bibinfo {volume} {82}},\
		\bibinfo {pages} {104401} (\bibinfo {year} {2019})}\BibitemShut {NoStop}%
	\bibitem [{\citenamefont {Mistakidis}\ \emph
		{et~al.}(2019{\natexlab{a}})\citenamefont {Mistakidis}, \citenamefont
		{Katsimiga}, \citenamefont {Koutentakis}, \citenamefont {Busch},\ and\
		\citenamefont {Schmelcher}}]{mistakidis2019quench}%
	\BibitemOpen
	\bibfield  {author} {\bibinfo {author} {\bibfnamefont {S.}~\bibnamefont
			{Mistakidis}}, \bibinfo {author} {\bibfnamefont {G.}~\bibnamefont
			{Katsimiga}}, \bibinfo {author} {\bibfnamefont {G.}~\bibnamefont
			{Koutentakis}}, \bibinfo {author} {\bibfnamefont {T.}~\bibnamefont {Busch}},\
		and\ \bibinfo {author} {\bibfnamefont {P.}~\bibnamefont {Schmelcher}},\
	}\href@noop {} {\bibfield  {journal} {\bibinfo  {journal} {Phys. Rev. Lett.}\
		}\textbf {\bibinfo {volume} {122}},\ \bibinfo {pages} {183001} (\bibinfo
		{year} {2019}{\natexlab{a}})}\BibitemShut {NoStop}%
	\bibitem [{\citenamefont {Mistakidis}\ \emph
		{et~al.}(2019{\natexlab{b}})\citenamefont {Mistakidis}, \citenamefont
		{Grusdt}, \citenamefont {Koutentakis},\ and\ \citenamefont
		{Schmelcher}}]{mistakidis2019dissipative}%
	\BibitemOpen
	\bibfield  {author} {\bibinfo {author} {\bibfnamefont {S.}~\bibnamefont
			{Mistakidis}}, \bibinfo {author} {\bibfnamefont {F.}~\bibnamefont {Grusdt}},
		\bibinfo {author} {\bibfnamefont {G.}~\bibnamefont {Koutentakis}},\ and\
		\bibinfo {author} {\bibfnamefont {P.}~\bibnamefont {Schmelcher}},\
	}\href@noop {} {\bibfield  {journal} {\bibinfo  {journal} {New J. Phys.}\
		}\textbf {\bibinfo {volume} {21}},\ \bibinfo {pages} {103026} (\bibinfo
		{year} {2019}{\natexlab{b}})}\BibitemShut {NoStop}%
	\bibitem [{\citenamefont {Mistakidis}\ \emph
		{et~al.}(2019{\natexlab{c}})\citenamefont {Mistakidis}, \citenamefont
		{Volosniev}, \citenamefont {Zinner},\ and\ \citenamefont
		{Schmelcher}}]{mistakidis2019effective}%
	\BibitemOpen
	\bibfield  {author} {\bibinfo {author} {\bibfnamefont {S.}~\bibnamefont
			{Mistakidis}}, \bibinfo {author} {\bibfnamefont {A.}~\bibnamefont
			{Volosniev}}, \bibinfo {author} {\bibfnamefont {N.}~\bibnamefont {Zinner}},\
		and\ \bibinfo {author} {\bibfnamefont {P.}~\bibnamefont {Schmelcher}},\
	}\href@noop {} {\bibfield  {journal} {\bibinfo  {journal} {Phys. Rev. A}\
		}\textbf {\bibinfo {volume} {100}},\ \bibinfo {pages} {013619} (\bibinfo
		{year} {2019}{\natexlab{c}})}\BibitemShut {NoStop}%
	\bibitem [{\citenamefont {Mistakidis}\ \emph {et~al.}(2020)\citenamefont
		{Mistakidis}, \citenamefont {Koutentakis}, \citenamefont {Katsimiga},
		\citenamefont {Busch},\ and\ \citenamefont
		{Schmelcher}}]{mistakidis2020many}%
	\BibitemOpen
	\bibfield  {author} {\bibinfo {author} {\bibfnamefont {S.}~\bibnamefont
			{Mistakidis}}, \bibinfo {author} {\bibfnamefont {G.}~\bibnamefont
			{Koutentakis}}, \bibinfo {author} {\bibfnamefont {G.}~\bibnamefont
			{Katsimiga}}, \bibinfo {author} {\bibfnamefont {T.}~\bibnamefont {Busch}},\
		and\ \bibinfo {author} {\bibfnamefont {P.}~\bibnamefont {Schmelcher}},\
	}\href@noop {} {\bibfield  {journal} {\bibinfo  {journal} {New J. Phys.}\
		}\textbf {\bibinfo {volume} {22}},\ \bibinfo {pages} {043007} (\bibinfo
		{year} {2020})}\BibitemShut {NoStop}%
	\bibitem [{\citenamefont {Schollw{\"o}ck}(2005)}]{schollwock2005density}%
	\BibitemOpen
	\bibfield  {author} {\bibinfo {author} {\bibfnamefont {U.}~\bibnamefont
			{Schollw{\"o}ck}},\ }\href@noop {} {\bibfield  {journal} {\bibinfo  {journal}
			{Rev. Mod. Phys.}\ }\textbf {\bibinfo {volume} {77}},\ \bibinfo {pages} {259}
		(\bibinfo {year} {2005})}\BibitemShut {NoStop}%
	\bibitem [{\citenamefont {Cao}\ \emph {et~al.}(2017)\citenamefont {Cao},
		\citenamefont {Bolsinger}, \citenamefont {Mistakidis}, \citenamefont
		{Koutentakis}, \citenamefont {Kr{\"o}nke}, \citenamefont {Schurer},\ and\
		\citenamefont {Schmelcher}}]{cao2017unified}%
	\BibitemOpen
	\bibfield  {author} {\bibinfo {author} {\bibfnamefont {L.}~\bibnamefont
			{Cao}}, \bibinfo {author} {\bibfnamefont {V.}~\bibnamefont {Bolsinger}},
		\bibinfo {author} {\bibfnamefont {S.}~\bibnamefont {Mistakidis}}, \bibinfo
		{author} {\bibfnamefont {G.}~\bibnamefont {Koutentakis}}, \bibinfo {author}
		{\bibfnamefont {S.}~\bibnamefont {Kr{\"o}nke}}, \bibinfo {author}
		{\bibfnamefont {J.}~\bibnamefont {Schurer}},\ and\ \bibinfo {author}
		{\bibfnamefont {P.}~\bibnamefont {Schmelcher}},\ }\href@noop {} {\bibfield
		{journal} {\bibinfo  {journal} {J. Chem. Phys}\ }\textbf {\bibinfo {volume}
			{147}},\ \bibinfo {pages} {044106} (\bibinfo {year} {2017})}\BibitemShut
	{NoStop}%
	\bibitem [{\citenamefont {Bentine}\ \emph {et~al.}(2017)\citenamefont
		{Bentine}, \citenamefont {Harte}, \citenamefont {Luksch}, \citenamefont
		{Barker}, \citenamefont {Mur-Petit}, \citenamefont {Yuen},\ and\
		\citenamefont {Foot}}]{bentine2017species}%
	\BibitemOpen
	\bibfield  {author} {\bibinfo {author} {\bibfnamefont {E.}~\bibnamefont
			{Bentine}}, \bibinfo {author} {\bibfnamefont {T.}~\bibnamefont {Harte}},
		\bibinfo {author} {\bibfnamefont {K.}~\bibnamefont {Luksch}}, \bibinfo
		{author} {\bibfnamefont {A.}~\bibnamefont {Barker}}, \bibinfo {author}
		{\bibfnamefont {J.}~\bibnamefont {Mur-Petit}}, \bibinfo {author}
		{\bibfnamefont {B.}~\bibnamefont {Yuen}},\ and\ \bibinfo {author}
		{\bibfnamefont {C.}~\bibnamefont {Foot}},\ }\href@noop {} {\bibfield
		{journal} {\bibinfo  {journal} {J. Phys. B}\ }\textbf {\bibinfo {volume}
			{50}},\ \bibinfo {pages} {094002} (\bibinfo {year} {2017})}\BibitemShut
	{NoStop}%
	\bibitem [{\citenamefont {Barker}\ \emph {et~al.}(2020)\citenamefont {Barker},
		\citenamefont {Sunami}, \citenamefont {Garrick}, \citenamefont {Beregi},
		\citenamefont {Luksch}, \citenamefont {Bentine},\ and\ \citenamefont
		{Foot}}]{barker2020realising}%
	\BibitemOpen
	\bibfield  {author} {\bibinfo {author} {\bibfnamefont {A.}~\bibnamefont
			{Barker}}, \bibinfo {author} {\bibfnamefont {S.}~\bibnamefont {Sunami}},
		\bibinfo {author} {\bibfnamefont {D.}~\bibnamefont {Garrick}}, \bibinfo
		{author} {\bibfnamefont {A.}~\bibnamefont {Beregi}}, \bibinfo {author}
		{\bibfnamefont {K.}~\bibnamefont {Luksch}}, \bibinfo {author} {\bibfnamefont
			{E.}~\bibnamefont {Bentine}},\ and\ \bibinfo {author} {\bibfnamefont
			{C.}~\bibnamefont {Foot}},\ }\href@noop {} {\bibfield  {journal} {\bibinfo
			{journal} {J. Phys. B}\ }\textbf {\bibinfo {volume} {53}},\ \bibinfo {pages}
		{155001} (\bibinfo {year} {2020})}\BibitemShut {NoStop}%
	\bibitem [{\citenamefont {Keiler}\ and\ \citenamefont
		{Schmelcher}(2019)}]{keiler2019interaction}%
	\BibitemOpen
	\bibfield  {author} {\bibinfo {author} {\bibfnamefont {K.}~\bibnamefont
			{Keiler}}\ and\ \bibinfo {author} {\bibfnamefont {P.}~\bibnamefont
			{Schmelcher}},\ }\href@noop {} {\bibfield  {journal} {\bibinfo  {journal}
			{Phys. Rev. A}\ }\textbf {\bibinfo {volume} {100}},\ \bibinfo {pages}
		{043616} (\bibinfo {year} {2019})}\BibitemShut {NoStop}%
	\bibitem [{\citenamefont {Pyzh}\ and\ \citenamefont
		{Schmelcher}(2020)}]{pyzh2020phase}%
	\BibitemOpen
	\bibfield  {author} {\bibinfo {author} {\bibfnamefont {M.}~\bibnamefont
			{Pyzh}}\ and\ \bibinfo {author} {\bibfnamefont {P.}~\bibnamefont
			{Schmelcher}},\ }\href@noop {} {\bibfield  {journal} {\bibinfo  {journal}
			{Phys. Rev. A}\ }\textbf {\bibinfo {volume} {102}},\ \bibinfo {pages}
		{023305} (\bibinfo {year} {2020})}\BibitemShut {NoStop}%
	\bibitem [{\citenamefont {Pyzh}\ \emph {et~al.}(2021)\citenamefont {Pyzh},
		\citenamefont {Keiler}, \citenamefont {Mistakidis},\ and\ \citenamefont
		{Schmelcher}}]{pyzh2021entangling}%
	\BibitemOpen
	\bibfield  {author} {\bibinfo {author} {\bibfnamefont {M.}~\bibnamefont
			{Pyzh}}, \bibinfo {author} {\bibfnamefont {K.}~\bibnamefont {Keiler}},
		\bibinfo {author} {\bibfnamefont {S.~I.}\ \bibnamefont {Mistakidis}},\ and\
		\bibinfo {author} {\bibfnamefont {P.}~\bibnamefont {Schmelcher}},\
	}\href@noop {} {\bibfield  {journal} {\bibinfo  {journal} {Entropy}\ }\textbf
		{\bibinfo {volume} {23}},\ \bibinfo {pages} {290} (\bibinfo {year}
		{2021})}\BibitemShut {NoStop}%
	\bibitem [{\citenamefont {Lee}\ \emph {et~al.}(1953)\citenamefont {Lee},
		\citenamefont {Low},\ and\ \citenamefont {Pines}}]{lee1953motion}%
	\BibitemOpen
	\bibfield  {author} {\bibinfo {author} {\bibfnamefont {T.}~\bibnamefont
			{Lee}}, \bibinfo {author} {\bibfnamefont {F.}~\bibnamefont {Low}},\ and\
		\bibinfo {author} {\bibfnamefont {D.}~\bibnamefont {Pines}},\ }\href@noop {}
	{\bibfield  {journal} {\bibinfo  {journal} {Phys. Rev.}\ }\textbf {\bibinfo
			{volume} {90}},\ \bibinfo {pages} {297} (\bibinfo {year} {1953})}\BibitemShut
	{NoStop}%
	\bibitem [{\citenamefont {Johnson}\ \emph {et~al.}(2012)\citenamefont
		{Johnson}, \citenamefont {Bruderer}, \citenamefont {Cai}, \citenamefont
		{Clark}, \citenamefont {Bao},\ and\ \citenamefont
		{Jaksch}}]{johnson2012breathing}%
	\BibitemOpen
	\bibfield  {author} {\bibinfo {author} {\bibfnamefont {T.~H.}\ \bibnamefont
			{Johnson}}, \bibinfo {author} {\bibfnamefont {M.}~\bibnamefont {Bruderer}},
		\bibinfo {author} {\bibfnamefont {Y.}~\bibnamefont {Cai}}, \bibinfo {author}
		{\bibfnamefont {S.~R.}\ \bibnamefont {Clark}}, \bibinfo {author}
		{\bibfnamefont {W.}~\bibnamefont {Bao}},\ and\ \bibinfo {author}
		{\bibfnamefont {D.}~\bibnamefont {Jaksch}},\ }\href@noop {} {\bibfield
		{journal} {\bibinfo  {journal} {EPL}\ }\textbf {\bibinfo {volume} {98}},\
		\bibinfo {pages} {26001} (\bibinfo {year} {2012})}\BibitemShut {NoStop}%
	\bibitem [{\citenamefont {Moritz}\ \emph {et~al.}(2003)\citenamefont {Moritz},
		\citenamefont {St{\"o}ferle}, \citenamefont {K{\"o}hl},\ and\ \citenamefont
		{Esslinger}}]{moritz2003exciting}%
	\BibitemOpen
	\bibfield  {author} {\bibinfo {author} {\bibfnamefont {H.}~\bibnamefont
			{Moritz}}, \bibinfo {author} {\bibfnamefont {T.}~\bibnamefont
			{St{\"o}ferle}}, \bibinfo {author} {\bibfnamefont {M.}~\bibnamefont
			{K{\"o}hl}},\ and\ \bibinfo {author} {\bibfnamefont {T.}~\bibnamefont
			{Esslinger}},\ }\href@noop {} {\bibfield  {journal} {\bibinfo  {journal}
			{Phys. Rev. Lett.}\ }\textbf {\bibinfo {volume} {91}},\ \bibinfo {pages}
		{250402} (\bibinfo {year} {2003})}\BibitemShut {NoStop}%
	\bibitem [{\citenamefont {St{\"o}ferle}\ \emph {et~al.}(2004)\citenamefont
		{St{\"o}ferle}, \citenamefont {Moritz}, \citenamefont {Schori}, \citenamefont
		{K{\"o}hl},\ and\ \citenamefont {Esslinger}}]{stoferle2004transition}%
	\BibitemOpen
	\bibfield  {author} {\bibinfo {author} {\bibfnamefont {T.}~\bibnamefont
			{St{\"o}ferle}}, \bibinfo {author} {\bibfnamefont {H.}~\bibnamefont
			{Moritz}}, \bibinfo {author} {\bibfnamefont {C.}~\bibnamefont {Schori}},
		\bibinfo {author} {\bibfnamefont {M.}~\bibnamefont {K{\"o}hl}},\ and\
		\bibinfo {author} {\bibfnamefont {T.}~\bibnamefont {Esslinger}},\ }\href@noop
	{} {\bibfield  {journal} {\bibinfo  {journal} {Phys. Rev. Lett.}\ }\textbf
		{\bibinfo {volume} {92}},\ \bibinfo {pages} {130403} (\bibinfo {year}
		{2004})}\BibitemShut {NoStop}%
	\bibitem [{\citenamefont {Haller}\ \emph {et~al.}(2009)\citenamefont {Haller},
		\citenamefont {Gustavsson}, \citenamefont {Mark}, \citenamefont {Danzl},
		\citenamefont {Hart}, \citenamefont {Pupillo},\ and\ \citenamefont
		{N{\"a}gerl}}]{haller2009realization}%
	\BibitemOpen
	\bibfield  {author} {\bibinfo {author} {\bibfnamefont {E.}~\bibnamefont
			{Haller}}, \bibinfo {author} {\bibfnamefont {M.}~\bibnamefont {Gustavsson}},
		\bibinfo {author} {\bibfnamefont {M.~J.}\ \bibnamefont {Mark}}, \bibinfo
		{author} {\bibfnamefont {J.~G.}\ \bibnamefont {Danzl}}, \bibinfo {author}
		{\bibfnamefont {R.}~\bibnamefont {Hart}}, \bibinfo {author} {\bibfnamefont
			{G.}~\bibnamefont {Pupillo}},\ and\ \bibinfo {author} {\bibfnamefont {H.-C.}\
			\bibnamefont {N{\"a}gerl}},\ }\href@noop {} {\bibfield  {journal} {\bibinfo
			{journal} {Science}\ }\textbf {\bibinfo {volume} {325}},\ \bibinfo {pages}
		{1224} (\bibinfo {year} {2009})}\BibitemShut {NoStop}%
	\bibitem [{\citenamefont {Fang}\ \emph {et~al.}(2014)\citenamefont {Fang},
		\citenamefont {Carleo}, \citenamefont {Johnson},\ and\ \citenamefont
		{Bouchoule}}]{fang2014quench}%
	\BibitemOpen
	\bibfield  {author} {\bibinfo {author} {\bibfnamefont {B.}~\bibnamefont
			{Fang}}, \bibinfo {author} {\bibfnamefont {G.}~\bibnamefont {Carleo}},
		\bibinfo {author} {\bibfnamefont {A.}~\bibnamefont {Johnson}},\ and\ \bibinfo
		{author} {\bibfnamefont {I.}~\bibnamefont {Bouchoule}},\ }\href@noop {}
	{\bibfield  {journal} {\bibinfo  {journal} {Phys. Rev. Lett.}\ }\textbf
		{\bibinfo {volume} {113}},\ \bibinfo {pages} {035301} (\bibinfo {year}
		{2014})}\BibitemShut {NoStop}%
	\bibitem [{\citenamefont {Abraham}\ and\ \citenamefont
		{Bonitz}(2014)}]{abraham2014quantum}%
	\BibitemOpen
	\bibfield  {author} {\bibinfo {author} {\bibfnamefont {J.}~\bibnamefont
			{Abraham}}\ and\ \bibinfo {author} {\bibfnamefont {M.}~\bibnamefont
			{Bonitz}},\ }\href@noop {} {\bibfield  {journal} {\bibinfo  {journal}
			{Contrib. to Plasma Phys.}\ }\textbf {\bibinfo {volume} {54}},\ \bibinfo
		{pages} {27} (\bibinfo {year} {2014})}\BibitemShut {NoStop}%
	\bibitem [{\citenamefont {McDonald}\ \emph {et~al.}(2013)\citenamefont
		{McDonald}, \citenamefont {Orlando}, \citenamefont {Abraham}, \citenamefont
		{Hochstuhl}, \citenamefont {Bonitz},\ and\ \citenamefont
		{Brabec}}]{mcdonald2013theory}%
	\BibitemOpen
	\bibfield  {author} {\bibinfo {author} {\bibfnamefont {C.}~\bibnamefont
			{McDonald}}, \bibinfo {author} {\bibfnamefont {G.}~\bibnamefont {Orlando}},
		\bibinfo {author} {\bibfnamefont {J.}~\bibnamefont {Abraham}}, \bibinfo
		{author} {\bibfnamefont {D.}~\bibnamefont {Hochstuhl}}, \bibinfo {author}
		{\bibfnamefont {M.}~\bibnamefont {Bonitz}},\ and\ \bibinfo {author}
		{\bibfnamefont {T.}~\bibnamefont {Brabec}},\ }\href@noop {} {\bibfield
		{journal} {\bibinfo  {journal} {Phys. Rev. Lett.}\ }\textbf {\bibinfo
			{volume} {111}},\ \bibinfo {pages} {256801} (\bibinfo {year}
		{2013})}\BibitemShut {NoStop}%
	\bibitem [{\citenamefont {Astrakharchik}\ \emph {et~al.}(2005)\citenamefont
		{Astrakharchik}, \citenamefont {Combescot}, \citenamefont {Leyronas},\ and\
		\citenamefont {Stringari}}]{astrakharchik2005equation}%
	\BibitemOpen
	\bibfield  {author} {\bibinfo {author} {\bibfnamefont {G.}~\bibnamefont
			{Astrakharchik}}, \bibinfo {author} {\bibfnamefont {R.}~\bibnamefont
			{Combescot}}, \bibinfo {author} {\bibfnamefont {X.}~\bibnamefont
			{Leyronas}},\ and\ \bibinfo {author} {\bibfnamefont {S.}~\bibnamefont
			{Stringari}},\ }\href@noop {} {\bibfield  {journal} {\bibinfo  {journal}
			{Phys. Rev. Lett.}\ }\textbf {\bibinfo {volume} {95}},\ \bibinfo {pages}
		{030404} (\bibinfo {year} {2005})}\BibitemShut {NoStop}%
	\bibitem [{\citenamefont {Altmeyer}\ \emph {et~al.}(2007)\citenamefont
		{Altmeyer}, \citenamefont {Riedl}, \citenamefont {Kohstall}, \citenamefont
		{Wright}, \citenamefont {Geursen}, \citenamefont {Bartenstein}, \citenamefont
		{Chin}, \citenamefont {Denschlag},\ and\ \citenamefont
		{Grimm}}]{altmeyer2007precision}%
	\BibitemOpen
	\bibfield  {author} {\bibinfo {author} {\bibfnamefont {A.}~\bibnamefont
			{Altmeyer}}, \bibinfo {author} {\bibfnamefont {S.}~\bibnamefont {Riedl}},
		\bibinfo {author} {\bibfnamefont {C.}~\bibnamefont {Kohstall}}, \bibinfo
		{author} {\bibfnamefont {M.}~\bibnamefont {Wright}}, \bibinfo {author}
		{\bibfnamefont {R.}~\bibnamefont {Geursen}}, \bibinfo {author} {\bibfnamefont
			{M.}~\bibnamefont {Bartenstein}}, \bibinfo {author} {\bibfnamefont
			{C.}~\bibnamefont {Chin}}, \bibinfo {author} {\bibfnamefont {J.~H.}\
			\bibnamefont {Denschlag}},\ and\ \bibinfo {author} {\bibfnamefont
			{R.}~\bibnamefont {Grimm}},\ }\href@noop {} {\bibfield  {journal} {\bibinfo
			{journal} {Phys. Rev. Lett.}\ }\textbf {\bibinfo {volume} {98}},\ \bibinfo
		{pages} {040401} (\bibinfo {year} {2007})}\BibitemShut {NoStop}%
	\bibitem [{\citenamefont {Andrade}\ \emph {et~al.}(2012)\citenamefont
		{Andrade}, \citenamefont {Sanders},\ and\ \citenamefont
		{Aspuru-Guzik}}]{andrade2012application}%
	\BibitemOpen
	\bibfield  {author} {\bibinfo {author} {\bibfnamefont {X.}~\bibnamefont
			{Andrade}}, \bibinfo {author} {\bibfnamefont {J.~N.}\ \bibnamefont
			{Sanders}},\ and\ \bibinfo {author} {\bibfnamefont {A.}~\bibnamefont
			{Aspuru-Guzik}},\ }\href@noop {} {\bibfield  {journal} {\bibinfo  {journal}
			{PNAS}\ }\textbf {\bibinfo {volume} {109}},\ \bibinfo {pages} {13928}
		(\bibinfo {year} {2012})}\BibitemShut {NoStop}%
	\bibitem [{\citenamefont {Pyzh}\ \emph {et~al.}(2018)\citenamefont {Pyzh},
		\citenamefont {Kr{\"o}nke}, \citenamefont {Weitenberg},\ and\ \citenamefont
		{Schmelcher}}]{pyzh2018spectral}%
	\BibitemOpen
	\bibfield  {author} {\bibinfo {author} {\bibfnamefont {M.}~\bibnamefont
			{Pyzh}}, \bibinfo {author} {\bibfnamefont {S.}~\bibnamefont {Kr{\"o}nke}},
		\bibinfo {author} {\bibfnamefont {C.}~\bibnamefont {Weitenberg}},\ and\
		\bibinfo {author} {\bibfnamefont {P.}~\bibnamefont {Schmelcher}},\
	}\href@noop {} {\bibfield  {journal} {\bibinfo  {journal} {New J. Phys.}\
		}\textbf {\bibinfo {volume} {20}},\ \bibinfo {pages} {015006} (\bibinfo
		{year} {2018})}\BibitemShut {NoStop}%
	\bibitem [{\citenamefont {Cao}\ \emph {et~al.}(2013)\citenamefont {Cao},
		\citenamefont {Kr{\"o}nke}, \citenamefont {Vendrell},\ and\ \citenamefont
		{Schmelcher}}]{cao2013multi}%
	\BibitemOpen
	\bibfield  {author} {\bibinfo {author} {\bibfnamefont {L.}~\bibnamefont
			{Cao}}, \bibinfo {author} {\bibfnamefont {S.}~\bibnamefont {Kr{\"o}nke}},
		\bibinfo {author} {\bibfnamefont {O.}~\bibnamefont {Vendrell}},\ and\
		\bibinfo {author} {\bibfnamefont {P.}~\bibnamefont {Schmelcher}},\
	}\href@noop {} {\bibfield  {journal} {\bibinfo  {journal} {J. Chem. Phys}\
		}\textbf {\bibinfo {volume} {139}},\ \bibinfo {pages} {134103} (\bibinfo
		{year} {2013})}\BibitemShut {NoStop}%
	\bibitem [{\citenamefont {Kr{\"o}nke}\ \emph {et~al.}(2013)\citenamefont
		{Kr{\"o}nke}, \citenamefont {Cao}, \citenamefont {Vendrell},\ and\
		\citenamefont {Schmelcher}}]{kronke2013non}%
	\BibitemOpen
	\bibfield  {author} {\bibinfo {author} {\bibfnamefont {S.}~\bibnamefont
			{Kr{\"o}nke}}, \bibinfo {author} {\bibfnamefont {L.}~\bibnamefont {Cao}},
		\bibinfo {author} {\bibfnamefont {O.}~\bibnamefont {Vendrell}},\ and\
		\bibinfo {author} {\bibfnamefont {P.}~\bibnamefont {Schmelcher}},\
	}\href@noop {} {\bibfield  {journal} {\bibinfo  {journal} {New J. Phys.}\
		}\textbf {\bibinfo {volume} {15}},\ \bibinfo {pages} {063018} (\bibinfo
		{year} {2013})}\BibitemShut {NoStop}%
	\bibitem [{\citenamefont {Light}\ \emph {et~al.}(1985)\citenamefont {Light},
		\citenamefont {Hamilton},\ and\ \citenamefont {Lill}}]{light1985generalized}%
	\BibitemOpen
	\bibfield  {author} {\bibinfo {author} {\bibfnamefont {J.}~\bibnamefont
			{Light}}, \bibinfo {author} {\bibfnamefont {I.}~\bibnamefont {Hamilton}},\
		and\ \bibinfo {author} {\bibfnamefont {J.}~\bibnamefont {Lill}},\ }\href@noop
	{} {\bibfield  {journal} {\bibinfo  {journal} {J. Chem. Phys}\ }\textbf
		{\bibinfo {volume} {82}},\ \bibinfo {pages} {1400} (\bibinfo {year}
		{1985})}\BibitemShut {NoStop}%
	\bibitem [{\citenamefont {Raab}(2000)}]{raab2000dirac}%
	\BibitemOpen
	\bibfield  {author} {\bibinfo {author} {\bibfnamefont {A.}~\bibnamefont
			{Raab}},\ }\href@noop {} {\bibfield  {journal} {\bibinfo  {journal} {Chem.
				Phys. Lett.}\ }\textbf {\bibinfo {volume} {319}},\ \bibinfo {pages} {674}
		(\bibinfo {year} {2000})}\BibitemShut {NoStop}%
	\bibitem [{\citenamefont {Van Den~Berg}\ and\ \citenamefont
		{Friedlander}(2009)}]{van2009probing}%
	\BibitemOpen
	\bibfield  {author} {\bibinfo {author} {\bibfnamefont {E.}~\bibnamefont {Van
				Den~Berg}}\ and\ \bibinfo {author} {\bibfnamefont {M.~P.}\ \bibnamefont
			{Friedlander}},\ }\href@noop {} {\bibfield  {journal} {\bibinfo  {journal}
			{SIAM J. Sci. Comput.}\ }\textbf {\bibinfo {volume} {31}},\ \bibinfo {pages}
		{890} (\bibinfo {year} {2009})}\BibitemShut {NoStop}%
	\bibitem [{\citenamefont {Loris}(2008)}]{loris2008l1packv2}%
	\BibitemOpen
	\bibfield  {author} {\bibinfo {author} {\bibfnamefont {I.}~\bibnamefont
			{Loris}},\ }\href@noop {} {\bibfield  {journal} {\bibinfo  {journal} {Comput.
				Phys. Commun}\ }\textbf {\bibinfo {volume} {179}},\ \bibinfo {pages} {895}
		(\bibinfo {year} {2008})}\BibitemShut {NoStop}%
	\bibitem [{\citenamefont {Schmitz}\ \emph {et~al.}(2013)\citenamefont
		{Schmitz}, \citenamefont {Kr{\"o}nke}, \citenamefont {Cao},\ and\
		\citenamefont {Schmelcher}}]{schmitz2013quantum}%
	\BibitemOpen
	\bibfield  {author} {\bibinfo {author} {\bibfnamefont {R.}~\bibnamefont
			{Schmitz}}, \bibinfo {author} {\bibfnamefont {S.}~\bibnamefont {Kr{\"o}nke}},
		\bibinfo {author} {\bibfnamefont {L.}~\bibnamefont {Cao}},\ and\ \bibinfo
		{author} {\bibfnamefont {P.}~\bibnamefont {Schmelcher}},\ }\href@noop {}
	{\bibfield  {journal} {\bibinfo  {journal} {Phys. Rev. A}\ }\textbf {\bibinfo
			{volume} {88}},\ \bibinfo {pages} {043601} (\bibinfo {year}
		{2013})}\BibitemShut {NoStop}%
\end{thebibliography}

\end{document}